%% file: ver9.tex
\documentclass[12pt,iop,twocolumn,apj]{emulateapj}
\usepackage[latin1]{inputenc}
\usepackage{amsmath}
\usepackage{amsfonts}
\usepackage{amssymb}
\usepackage{graphicx}
\usepackage{graphics}
\usepackage{threeparttable}
\shorttitle{Feathering Instability of Spiral Arms.  I.  Formulation of the Problem}
\shortauthors{Lee and Shu}

\input{symbols.tex}
\begin{document}
\title{FEATHERING INSTABILITY OF SPIRAL ARMS. \\
I: FORMULATION OF THE PROBLEM}
\author{Wing-Kit Lee$^1$ and Frank H. Shu $^{1,2}$}
\affil{$^1$Center for Astrophysics and Space Sciences \\
University of California, San Diego, La Jolla, CA 92093-0424}
\affil{$^2$Institute of Astronomy and Astrophysics, Academia Sinica, Taipei 115, Taiwan, Republic of China}
\email{wklee@ucsd.edu}

\begin{abstract}
In this paper we study the feathering substructures along spiral arms by considering the perturbational gas response to a spiral shock. Feathers are density fluctuations that jut out from the spiral arm to the inter-arm region at pitch angles given by the quantum numbers of the doubly-periodic structure. In a localized asymptotic approximation, related to the shearing sheet except that the inhomogeneities occur in space rather than in time, we derive the linearized perturbation equations for a razor-thin disk with turbulent interstellar gas, frozen-in magnetic field, and gaseous self-gravity.  Apart from the modal quantum numbers, the individual normal modes of the system depend on seven dimensionless quantities that characterize the underlying time-independent axisymmetric state plus its steady, nonlinear, two-armed spiral-shock (TASS) response to a hypothesized background density-wave supported by the disk stars of the galaxy.  We show that some of these normal modes have positive growth rates.  Their over-density contours in the post-shock region are very reminiscent of observed feathering substructures in full magnetohydrodynamic (MHD) simulations. The feathering substructures are parasitic instabilities intrinsic to the system; thus, their study not only provides potential diagnostics for important parameters that characterize the interstellar medium of external galaxies, but also yields a deeper understanding of the basic mechanism that drives the formation of the giant molecular clouds (GMCs) and the OB stars that outline observed grand-design spirals.

\end{abstract}

\section{INTRODUCTION}
Spiral structures in nearby galaxies have fascinated astronomers since Lord Rosse's observations of M51 in 1845. The underpinning for a theoretical understanding of the phenomenon in terms of density waves has existed about 50 years (see \citealt{1964ApJ...140..646L} and references therein). Improved imaging technology and techniques reveal many substructures associated with the spiral arms.  Here, we focus on the quasi-regularly spaced density fluctuations identified in the literature as feathers \citep{1970IAUS...38...26L} or spurs \citep{1980ApJ...242..528E}. Observationally, the feathers are extinction substructure commonly found in the optical band images among spiral galaxies, e.g., \citet{2006ApJ...650..818L}. For example, a Hubble Heritage image of M51 \citep{Scoville2001press}, shows many feathers (i.e., darkened dust lane in the optical image) projected into the inter-arm region from the primary dust lane. There are also examples showing the feathers in infrared (e.g., the $8\mu \rm m$ image of M81 from \textit{Spitzers Space Telescope}) or sub-millimeter wavelengths, such as detection of CO emission in M51 feathers \citep{2008ApJ...689..148C}. Therefore, the relationship between the feathers and the underlying interstellar medium (molecular and atomic gas, dust, magnetic field, etc.) may hold the key to an understanding of the formation of the GMCs and OB stars that delineate the arms of spiral galaxies.

There are two points of view regarding the background structures of spiral galaxies.  The first is the hypothesis of quasi-stationary spiral structure (QSSS) that attributes the origin of spiral structure to the normal modes of the disk stars of a flattened galaxy. This point of view seems consistent with the observational finding that spiral galaxies, which look fragmentary, multi-armed, and even flocculent at optical or blue wavelengths, nevertheless have, in $2.1\mu \rm m$ images, the grand-design two-armed spiral structure (TASS) that underlies the QSSS hypothesis (e.g., \citealt{1991Natur.353...48B}; \citealt{1994A&A...288..365B}; \citealt*{1996Natur.381..674B}). The second comes from numerical simulations
that show nonlinear effects saturating the growth of unstable normal modes \citep[e.g.,][]{2012ApJ...751...44S} lead to spiral patterns that are locally transient. These are dichotomies of long standing that we do not address in the present paper, which focuses on the substructures that arise from the response of the self-gravitating and magnetized interstellar medium even to the steady forcing associated with the classic QSSS hypothesis.

Theoretical understanding of the substructures is also confused. Explanations encompass both irregular causes such as swing-amplified shearing instabilities \citep[e.g.,][]{1965MNRAS.130..125G}, and regular causes such as gravitational instabilities \citep[e.g.,][]{B88,KO2002} initiated by a TASS pattern \citep[][]{Roberts1969a,1970ApJ...161..887R}.   Another possibility is that spurs arise as response of the disk stars to over-dense regions like GMCs \citep[e.g.,][]{1966ApJ...146..810J, 2012arXiv1204.0513D}.  The last possibility will lead, however, to spurs with characteristic inclinations that co-rotate with the local material velocity of the GMCS, which would not have an obvious correlation with the spiral pattern of the older disk stars. Elmegreen's (1980) conclusion that spurs have characteristic inclinations that correlate with {\it I} band images of the older disk stellar population suggests that feathering is best described as a long-lived phenomenon, intimately connected with the underlying spiral structure of disk galaxies.  We adopt this hypothesis for the analysis of the present paper, and do not speculate on the changes necessary if spiral patterns are short-lived with spiral arms persistent only in a statistical sense.

The QSSS hypothesis implicitly underlies many numerical simulations in recent years on this subject \citep[e.g.,][]{KO2002,2003ApJ...596..220C,2006ApJ...646..213K,2006ApJ...647..997S,2006MNRAS.367..873D,2008MNRAS.391..844D}. These sophisticated simulations include MHD, self-gravity, ISM phases, etc. They provide a detailed time evolution of how GMCs can be formed by the fragmentation and agglomeration of interstellar gas by local Jeans instability. However, due to computational limitations, the behavior of the system is followed for only a few orbital times. Also, as we shall see, given that seven dimensionless numbers form the irreducible set that characterizes the instability of the system, a comprehensive survey of parameter space by numerical simulation is clearly out of the question for the foreseeable future.  On the other hand, most theoretical linearized-stability analyses along the same line of thought include restrictive assumptions such as an arbitrary background profile, and/or a shearing-sheet approach, and/or a lack of gaseous self-gravity and/or magnetic fields \citep[e.g.,][]{1996ApJ...467...87D}.  These simplifications compromise the applicability of the analysis if we wish ultimately to use the theory as a diagnostic of the physical conditions in real systems.  Our aim here is to rectify these shortcomings.

In this paper we formulate and solve the basic equations that govern the formation of feathers through the instability of a galactic spiral shock when the roles of gaseous self-gravity and magnetic field are included within the original TASS framework of \cite{Roberts1969a}.  We work in the frame that corotates with pattern speed $\Omega_{\rm p}$ of the spiral gravitational field of the background stellar disk, in which the TASS pattern is independent of time $t$ and asymptotically one-dimensional (i.e., variations only in the direction perpendicular to spiral arms). By transforming the governing nonlinear equations to a spiral coordinate system ($\eta,\xi)$ with $\eta$ varying perpendicular to spiral arms, and $\xi$ along them, we write down the asymptotic equations that govern nonlinear behavior in which the underlying TASS pattern varies only in $\eta$ but the parasitic perturbations above the TASS pattern can vary in all three variables $(\eta,\xi,t)$.  The self-consistency of the asymptotic approximation then requires us to impose that single-valued perturbations are doubly periodic in $(\eta,\xi)$ when we linearize in the amplitude of the perturbations relative to the TASS state. This double-periodicity is characterized by two integers (quantum numbers): $m$ = the number of stellar spiral arms in a complete circle around the galaxy with $m$ assumed to equal to 2 in practice, and $l$ = the number of feathers as we go along a spiral arm that would take us to the next spiral arm (half-way circumferentially around the galaxy if $m = 2$) if we were to go instead in the direction perpendicular to a spiral arm.

Our calculation on the TASS part of the problem differs from the original Roberts work in that we include frozen-in magnetic fields \citep[as did][]{1970ApJ...161..887R} and gaseous self-gravity (Ostriker and Kim's 2002 analysis included only the self-gravity of the feathering perturbations, and not its effect on the underlying TASS state).  When we also include the effect of turbulent motions of the interstellar gas, modelled as a ``logatropic" gas (pressure $P$ proportional to the logarithm of the density $\rho$), there are seven dimensionless, irreducible, numbers that characterize the TASS state: (1) the ratio of the circular frequency to the epicyclic frequency $\Omega/\kappa$; (2) the sine of the inclination of the stellar spiral arms, $\sin i$; (3) the dimensionless Doppler-shifted frequency at which gas rotating at its circular angular speed $\Omega$ meets $m$ stellar spiral arms that each rotate at angular speed $\Omega_{\rm p}$, $-\nu = m(\Omega-\Omega_{\rm p})/\kappa$; (4) the amplitude of the stellar spiral gravitational field as a fraction of the axisymmetric radial gravitational field, $F$; (5) the  dimensionless measure of the gas surface density, $\alpha$; (6) the dimensionless measure of the mean gas turbulent speed, $x_{\rm t}$; and (7) the dimensionless measure of the mean Alfv\'en speed of the magnetized interstellar medium, $x_{\rm A}$.

The plan of the paper is as follows.  In \S \ref{basiceqs}, we write down the basic MHD equations in the spiral coordinates.  In \S \ref{spiralshock},  we obtain by the shooting method the 1-D nonlinear TASS solution in $\eta$ (across the spiral arm) modified from the Roberts-style analysis by the inclusion of magnetic fields and gaseous self-gravity.  In \S \ref{featpert}, we derive the equations that govern linearized, time-dependent, 2-D perturbations on top of the background TASS pattern. Because the basic reference state depends only on $\eta$ and not $\xi$ nor $t$, the linearized perturbations can be taken to be oscillatory (with complex frequency $\omega_{\rm R}+i\omega_{\rm I}$) in time $t$ and (with real dimensionless wavenumber $l$) in the spatial dimension $\xi$, but with dependences on the spatial dimension $\eta$ that satisfy ordinary differential equations.  Generality requires us to consider oscillatory perturbations in the position of the shock.  When the appropriate jump conditions (due to the corrugation of the shock front) are imposed on top of the condition of double-periodicity, we obtain the real and imaginary parts of the perturbation frequency, $\omega_{\rm R}$ and $\omega_{\rm I}$, as eigenvalues of the problem when the quantum numbers $m$ and $l$ are specified, together with the numerical values of $\Omega/\kappa$, $\nu$, $\sin i$, $F$, $\alpha$, $x_{\rm t}$, and $x_{\rm A}$. In \S \ref{results} we give a sample result.  In \S \ref{discussion} we discuss the physical meaning of the result and give our conclusions.

\section{BASIC EQUATIONS AND GEOMETRY}
\label{basiceqs}
We first write down the basic equations for the problem from the two-dimensional, time-dependent, ideal MHD equations in the rotating frame. We identify the axisymmetric, time-independent solution as the zeroth order state. We then introduce the tight-winding spiral arm approximation and obtain the $1^{st}$ order (in term of $\sin i$) nonlinear TASS state in the sense of \cite{Roberts1969a}, modified for gaseous self-gravity and the presence of frozen-in magnetic fields.
\subsection{Basic Equations}
In cylindrical polar coordinates $(\varpi, \varphi, z)$, we denote $\Sigma$, $u_\varpi$ and $u_\varphi$ as, respectively, the gas surface density, $\varpi$- and $\varphi$-components of the fluid velocity in a razor-thin flat disk. The continuity and momentum equations, in a rotating frame with angular rate of the spiral pattern, $\Omega_p$, can be written as,
\begin{equation}
\label{aa0}
\pd{\Sigma}{t}+\frac{1}{\varpi}\pd{}{\varpi}(\varpi\Sigma u_\varpi)+\frac{1}{\varpi}\pd{}{\varphi}(\Sigma u_\varphi)=0;
\end{equation}
\begin{align}
\label{aa1}\nonumber
\lefteqn{
\pd{u_\varpi}{t}+u_\varpi\pd{u_\varpi}{\varpi}+\frac{u_\varphi}{\varpi}\pd{u_\varpi}{\varphi}-\frac{u_\varphi^2}{\varpi}
}\\
=&\mathcal{F}_\varpi-\frac{1}{\Sigma}\pd{\Pi}{\varpi}-\pd{\mathcal{V}_{\rm eff}}{\varpi}+2\Omega_{\rm p} u_\varphi;
\end{align}

\begin{align}
\label{aa2}\nonumber
\lefteqn{
\pd{u_\varphi}{t}+u_\varpi\pd{u_\varphi}{\varpi}+\frac{u_\varphi}{\varpi}\pd{u_\varphi}{\varphi}+\frac{u_\varpi u_\varphi}{\varpi}
}\\
=&\mathcal{F}_\varphi-\frac{1}{\varpi\Sigma}\pd{\Pi}{\varphi}-\frac{1}{\varpi}\pd{\mathcal{V}_{\rm eff}}{\varphi}-2\Omega_{\rm p} u_\varpi;
\end{align}
where $\Pi$ is the vertically-integrated gas pressure, $\mathcal{F}_\varpi, \mathcal{F}_\varphi$ are the two horizontal components of the Lorentz force per unit mass. The last terms in equations (\ref{aa1}) and (\ref{aa2}) are the Coriolis accelerations associated with being in a frame of reference that rotates at angular speed $\Omega_{\rm p}$.  We write the effective potential as
\begin{equation}\label{aa3}
\mathcal{V}_{\rm eff}\equiv \mathcal{V}-\frac{1}{2}\Omega_{\rm p}^2 \varpi^2,
\end{equation}
where the the second term is the centrifugal contribution and the first term is the total gravitational potential of dark matter, stars, and gas evaluated in the plane of the disk $z=0$:
\begin{equation}\label{aa3b}
\mathcal{V}=\mathcal{V}_0(\varpi)+\mathcal{V}_*+(\varpi,\varphi) + \mathcal{V}_{\rm g}(\varpi,\varpi,t).
\end{equation}
The axisymmetric part $\mathcal{V}_0(\varpi)$ arises from the mass distribution of all three components (dark matter, stars, and interstellar gas).  It yields an angular speed $\Omega(\varpi)$ and an associated epicyclic frequency $\kappa (\varpi)$ defined by the gradient of the specific angular momentum, an expression that is also sometimes called the Rayleigh discriminant:
\begin{equation}\label{epicyclic}
\kappa^2 \equiv {1\over \varpi^3}{d\over d\varpi}\left[ \left(\varpi^2\Omega \right)^2\right].
\end{equation}
The role of $\Omega$ and $\kappa$ are well known in galactic dynamics, and their ratio $\Omega/\kappa$ is one of the fundamental dimensionless parameters in the current theory.

The quantity $\mathcal{V}_*$ is the (specified) spiral potential provided by the disk stars:
\begin{equation}\label{aa3c}
\mathcal{V}_* = -A(\varpi)\cos\left[m\varphi- \Phi(\varpi)\right],
\end{equation}
with $A(\varpi)$ and $\Phi(\varpi)$ being the amplitude and radial phase of the spiral gravitational potential, both of which are regarded here as given functions of galactocentric radius $\varpi$ from stellar density-wave theory.  In the convention of density wave theory, the radial wavenumber $k_\varpi \equiv \Phi^\prime (\varpi)$ is negative for trailing spiral waves.  The asymptotic (or WKBJ) approximation of spiral-density wave theory assumes a small tilt angle $i$ of the spiral arms, i.e., that $\tan i = m/|k_\varpi|\varpi$ is small compared to unity.  To justify the use of linear theory of a sinusoidal shape factor for the stellar spiral, the radial forcing amplitude of the stellar spiral arms, $|k_\varpi|A$, should be a small fraction of the axisymmetric gravitational acceleration:
\begin{equation}\label{bigF}
F \equiv {|k_\varpi|A \over \varpi\Omega^2}.
\end{equation}
A typical number quoted in the literature is $F = 5$ to 10$\%$. Although infrared images may indicate stronger fractions compared to the background stellar disk, especially in the outer disk, it should be recalled that the denominator in the definition of $F$ includes the force contribution from the dark matter halo. \citet*{SMR1973} (hereafter SMR) show, however, that the real measure of nonlinearity of the gaseous forcing is given by the combination,
\begin{equation}\label{smallf}
f \equiv \left({\Omega\over \kappa}\right)^2{mF\over \sin i},
\end{equation}
which is not a small parameter because the large factor $m/\sin i$ compensates for the small factor $F$.  A physical way of stating the same conclusion is that  the spiral gravitational field only needs to produce radial velocities comparable to the turbulent or Alfv\'en speeds to have large effects (e.g., shock waves) in the interstellar medium. The turbulent or Alfv\'en speeds are much smaller than the rotational velocities in giant spirals.

In this paper, we wish to study not only the effects of stellar forcing, but the enhancements produced by gaseous self-gravity.  In 3-D, the gaseous component of the gravitational potential, $\mathcal{V}_{\rm g}(\varpi, \varphi,z,t)$ is related to the gas surface density, $\Sigma$, by the Poisson equation for a razor-thin disk:
\begin{equation}\label{aa5}
{1 \over \varpi}{\partial \over \partial \varpi}\left( \varpi {\partial V_{\rm g}\over \partial \varpi}\right) + {1 \over \varpi^2}{\partial^2 V_{\rm g}\over \partial \varpi^2} +{\partial^2 V_{\rm g}\over \partial z^2} = 4\pi G\Sigma(\varpi,\varphi,t)\delta(z).
\end{equation}

In equations (\ref{aa1}) and (\ref{aa2}), the radial and tangential components of the Lorentz force per unit mass of the conducting fluid,
$\mathcal{F}_\varpi$ and $\mathcal{F}_\varphi$, are given by
\begin{equation}\label{aa5a}
\mathcal{F}_\varpi = -{z_0\over 2\pi \Sigma}{B_\varphi\over \varpi}\left[\pd{\left(\varpi B_\varphi\right)}{\varpi}-\pd{B_\varpi}{\varphi}\right],
\end{equation}
\begin{equation}\label{aa5b}
\mathcal {F}_\varphi = {z_0\over 2\pi\Sigma}{B_\varpi \over \varpi}\left[ \pd{\left(\varpi B_\varphi \right)}{\varpi}-\pd{B_\varpi}{\varphi}\right].
\end{equation}
where $z_0 \ll \varpi$ is the equivalent half-height of the gaseous disk over which the matter is realistically distributed.  In this paper, we implicitly assume that $z_0$ is a constant, but \cite{1973ApJ...179..755P} pointed out that this state of affairs would lead to an enhancement of synchrotron radiation behind spiral arms that is larger than was subsequently observed \citep*[e.g.,][]{1972A&A....17..468M}. 
\cite*{1974A&A....33...73M} proposed that magnetic buckling of the field and its subsequent inflation by cosmic rays \citep{1969SSRv....9..651P} could solve this difficulty. In the current analysis, we ignore this complication as well as the role of cosmic rays, but we warn that more accurate feathering analyses will need modification when the feather spacing becomes comparable to the disk thickness.

On the large scales of interest to the problem, the interstellar magnetic field can be assumed to satisfy the condition of field freezing for a planar magnetic field:
\begin{align}\label{aa6a}
\pd{B_\varpi}{t}+{1\over \varpi}\pd{}{\varphi}\left(B_\varpi u_\varphi -B_\varphi u_\varpi\right) = 0;\\ 
\pd{B_\varphi}{t}-\pd{}{\varpi}\left( B_\varpi u_\varphi - B_\varphi u_\varpi\right)=0.
\end{align}
Notice that $\varpi$ times the first equation followed by partial differentiation by $\varpi$ added to the partial differentiation of the second equation by $\varphi$ implies that the constraint of no magnetic monopoles,
\begin{equation}\label{aa6b}
{1\over \varpi}\pd{\left( \varpi B_\varpi \right)}{\varpi} +{1 \over \varpi} \pd{B_\varphi}{\varphi} = 0,
\end{equation}
holds for all time if it is satisfied initially.

Finally, to close our set of basic equations, we model the turbulent gas pressure with a logatropic equation of state \citep{1989ApJ...342..834L}:
\begin{equation}
\Pi_{\rm g} \equiv \Sigma_0 v_{\rm t0}^2 \ln\left(\frac{\Sigma}{\Sigma_0}\right),
\end{equation}
where $v_{\rm t0}$ is a characteristic turbulent speed. The square of the signal speed associated with this equation of state is given by
\begin{equation}\label{signal}
v_{\rm t}^2 \equiv {d\Pi_{\rm g}\over d\Sigma} = v_{\rm t0}^2 \left( {\Sigma_0 \over \Sigma} \right),
\end{equation}
which mimics the observed tendency for dense interstellar gas (e.g., molecular cloud complexes) to have lower turbulent speeds than rarified interstellar gas (e.g., H I clouds). The derivative of $\Pi_{\rm g}$ being positive and decreasing with increasing $\Sigma$ are more important properties of the logarithmic law than the formal feature of having a negative turbulent pressure in the regions of low surface density, because the formal pressure can contain an arbitrary addictive constant without having any physical effects on the analysis.

\subsection{Axisymmetric State}
We identify the axisymmetric quantities as the zeroth order reference state, and denote them by the subscript 0. With only circular velocities in the corotating frame, $u_\varphi = \varpi\left[\Omega(\varpi)-\Omega_{\rm p}\right]$, and toroidal magnetic fields, $B_\varphi = B_{\varphi 0}$, that depend on $\varpi$, the equation for radial force balance becomes
\begin{equation}\label{aa7a}
\varpi \Omega^2 = -{d{\mathcal V}_0\over d\varpi} -{1\over \Sigma_0}{d\Pi_0 \over d\varpi} - {z_0\over 2\pi \Sigma_0}{B_{\varphi 0}\over \varpi}{d\over d\varpi}\left(\varpi B_{\varphi 0}\right).
\end{equation} 

\subsection{TASS State}
In the corotating frame, the TASS state is also time-steady, so the field-freezing equation (\ref{aa6a}) can be satisfied, just as in the axisymmetric state, by assuming that the magnetic field is parallel to the vector velocity, which we can write in the form:
\begin{align}
\label{backgd7b}
B_\varpi(\varpi,\varphi)  &= b \Sigma(\varpi,\varphi) u_\varpi(\varpi,\varphi); \\
B_\varphi(\varpi,\varphi) &= b \Sigma(\varpi,\varphi) u_\varphi(\varpi,\varphi),
\end{align}
where the scalar factor of proportionally $b$ is chosen to be a constant in order to satisfy the condition of zero monopoles when the equation of continuity for the gas also holds (i.e., {\bf B} and $\Sigma {\bf u}$ both have zero two-dimensional divergence).  Because the fluid velocity is mostly circular even in the TASS flow, the $\varphi$-component of the magnetic field is much larger than its $\varpi$-component, except near corotation where $\Omega(\varpi) = \Omega_{\rm p}$.  Far from corotation, if we suppose the asymptotic approximation that the TASS flow produces radial variations that are large compared to tangential variations (or obtained by dividing perturbational quantities by $\varpi$), we may approximate the above expressions by
\begin{equation}\label{lorentz1}
\mathcal{F}_\varpi \simeq -\frac{z_0}{2\pi\Sigma}B_\varphi\partial_\varpi(B_\varpi)\simeq-v_{\rm A0}^2\pd{}{\varpi}\left(\frac{\Sigma}{\Sigma_0}\right),
\end{equation}
and,
\begin{equation}\label{lorentz2}
\mathcal{F}_\varphi \simeq \frac{z_0}{2\pi\Sigma}B_\varpi\partial_\varpi(B_\varphi)
\simeq v_{\rm A0}^2\frac{u_\varpi}{u_\varphi}\pd{}{\varpi}\left(\frac{\Sigma}{\Sigma_0}\right),
\end{equation}
where we define the square of the unperturbed Alf\'{v}en speed as
\begin{equation}\nonumber
v_{\rm A0}^2 \equiv \frac{B_{\varphi0}^2}{4\pi\Sigma_0/2z_0},
\end{equation}
and we have ignored the spatial variation of $\Sigma_0$ (axisymmetric part) in comparison with those of $\Sigma$.

In giant spiral galaxies, the squares of the characteristic turbulent and Alfv\'en speeds, $v^2_{\rm t0}$ and $v^2_{\rm A0}$ are small compared to the square of the flow velocity on the large scale, e.g., $\varpi^2\Omega^2$.  In these circumstances, the second and third terms on the right-hand-side of equation (\ref{aa7a}) are small in comparison to the first, and the rotation speed $\varpi\Omega(\varpi)$ depends mostly on the gravitational potential ${\mathcal V}_0(\varpi)$ of the axisymmetric distribution of dark plus ordinary matter, which we shall henceforth assume to be fixed.

The adoption of the logarithmic equation of state allows us to write $\Sigma^{-1} \nabla \Pi_{\rm g} = \nabla \mathcal{H}_{\rm g}$, where $\mathcal{H}_{\rm g}$ is the specific enthalpy of the turbulent gas:
\begin{equation}
\mathcal{H}_{\rm g} = -v^2_{\rm t0}\left(\frac{\Sigma_0}{\Sigma}\right).
\end{equation} 
Similarly, we may identify the ``specific enthalpy" associated with the dominant part of the magnetic ``pressure" $\Pi_{\rm m}=(v^2_{A0}/2)\Sigma_0^{-1}\Sigma^2$:
\begin{equation}
\mathcal{H}_{\rm m} = v^2_{\rm A0}\left(\frac{\Sigma}{\Sigma_0}\right).
\end{equation} 

\subsubsection*{Nonlinear perturbation}
We now return to the rest of our equations and assume that the actual situation is a combination of an axisymmetric, time-independent state plus a nonlinear TASS response and further feathering perturbations.  The subscript 1 in this section will refer to both the TASS response and the feathering instability, but in the Appendices we shall apply it only to the feathering perturbations and include the TASS response with the axisymmetric state (as unscripted variables when it will cause no confusion).  In ensuing sections, we avoid confusion by attaching a $\sim $ when we mean the perturbations due to the feathering instability alone.  Thus,
\begin{align}\label{nonlinearpert}
\nonumber
\Sigma &= \Sigma_0(\varpi) + \Sigma_1(\varpi,\varphi,t),\\ \nonumber
\mathcal{H} &= \mathcal{H}_0(\varpi) + \mathcal{H}_1(\varpi,\varphi,t),\\ \nonumber
u_\varpi &= u_{\varpi0}(\varpi) + u_1(\varpi,\varphi,t),\\ \nonumber
u_\varphi &= \varpi(\Omega-\Omega_p) + v_1(\varpi,\varphi,t),\\ \nonumber
\mathcal{F}_\varpi &= f_0(\varpi) + f_{\varpi1}(\varpi,\varphi,t),\\ \nonumber
\mathcal{F}_\varphi &= f_{\varphi1}(\varpi,\varphi,t),\\
\end{align}
where $\mathcal{H}$ is the enthalpy associated with the gas pressure and magnetic pressure. Note that only $v_1$ is small compared to its zeroth-order counterpart, with even the last approximation breaking down near corotation. Consistent with the approximation that the pressure gradients of the gas turbulent motions and magnetic fields contribute little to the axisymmetric force balance, we attribute their influence to the perturbations marked out by the subscript 1. Without linearization (because the TASS response is highly nonlinear), the substitution of equations (\ref{nonlinearpert}) yields the set:
\begin{align}
\nonumber
&\pd{u_1}{t}+u_1\pd{u_1}{\varpi}+\left(\Omega-\Omega_p+\frac{v_1}{\varpi}\right)\pd{u_1}{\varphi}-\frac{v_1^2}{\varpi}\\
=&2\Omega v_1-\pd{}{\varpi}(\mathcal{H}_{\rm g1}+\mathcal{H}_{\rm m1}+\mathcal{U}),
\end{align}
\begin{align}
\nonumber
&\pd{v_1}{t}+u_1\pd{v_1}{\varpi}+\left(\Omega-\Omega_p+\frac{v_1}{\varpi}\right)\pd{v_1}{\varphi}
+\frac{u_1 v_1}{\varpi}\\\nonumber
=&-\frac{\kappa^2}{2\Omega} u_1-\frac{1}{\varpi}\pd{}{\varphi}(\mathcal{H}_{\rm g1}+\mathcal{U})\\
&+\frac{u_1}{\varpi(\Omega-\Omega_{\rm p})+v_1}\pd{\mathcal{H}_{\rm m1}}{\varpi}.
\end{align}
In the above, $\mathcal{U} \equiv \mathcal{V}_*+\mathcal{V}_{\rm g}$ is the perturbation of the gravitational potential beyond the axisymmetric state.  Consistent with the approximation made above of ignoring the radius of curvature, the continuity equation becomes
\begin{equation}
\pd{\Sigma_1}{t}+\pd{}{\varpi}[(\Sigma_0+\Sigma_1)u_1]+\pd{}{\varphi}[\Sigma_1(\Omega-\Omega_{\rm p})]=0.
\end{equation}

\subsection{Spiral Coordinates and Asymptotic Approximation}
We follow \cite{Roberts1969a} and SMR in introducing the local orthogonal coordinates $(\eta,\xi)$ in the plane of the disk galaxy, where curves of $\xi$=constant and $\eta$=constant define, respectively, the directions perpendicular and parallel to a background stellar density waves with a locus of local gravitational potential minimum whenever the spiral phrase,
\begin{equation}
\eta (\varpi, \varphi) \equiv m\varphi - \Phi(\varpi)
\end{equation}
that enters in equation (\ref{aa3c}) equals zero or integer multiples of $2\pi$. But the function $\eta(\varpi,\varphi)$ can increase by $2\pi$ either by $\varphi$ increasing by $\pi$ (for $m = 2$) with $\varpi$ fixed (going around halfway the galaxy in a circle), or by $\Phi(\varpi)$ increasing by $2\pi$ with $\varphi$ fixed (going out radially until the next spiral arm).  We wish the solution to look the same in either case to lowest asymptotic order.  In a local treatment, where we approximate the inclination angle $i$ of the spiral arms with respect to the circular direction to be constant, then $\Phi(\varpi) = -(m/\tan i)\ln(\varpi/\varpi_0)$, which corresponds to the case when the stellar spiral arms are fitted by logarithmic spirals.

We now introduce the orthogonal spiral coordinates $\eta$ and $\xi$ used by SMR with $\hat{\mathbf{e}}_\eta \times \hat{\mathbf{e}}_\xi = \hat{\mathbf{e}}_z$ and
\begin{align}
d\eta = -\Phi'(\varpi)d\varpi+md\varphi = m\left(\frac{1}{\tan{i}}\frac{d\varpi}{\varpi}+d\varphi\right),\\
d\xi = \frac{m}{\tan i}\left[\frac{m}{\Phi'(\varpi)}\frac{d\varpi}{\varpi^2}+d\varphi\right]
= m\left(\frac{d\varpi}{\varpi}+\frac{1}{\tan i}d\varphi\right).
\end{align}
Note that if we move radially outward at fixed $\varpi$, $\xi$ will increase by $2\pi\cot i$ for the same increase in $\varpi$ that results in an increase of $2\pi$ for $\eta$. When we transform from $(\varpi,\varphi)$ to $(\eta, \xi)$ and draw rectangular boxes in $(\eta,\xi)$, the coordinate system is similar to the one defined in \cite{KO2002}, but there are two differences.  (1) We do the calculations in a standard Eulerian manner, without mixing time and space coordinates as in the ``shearing sheet" treatment.  (2) The ratio of the axes are depicted in their correct geometric proportions, determined by the spiral pitch angle $i$ (see Fig. 2). 

\begin{figure}[!ht]
\begin{center}
\includegraphics[scale=0.45]{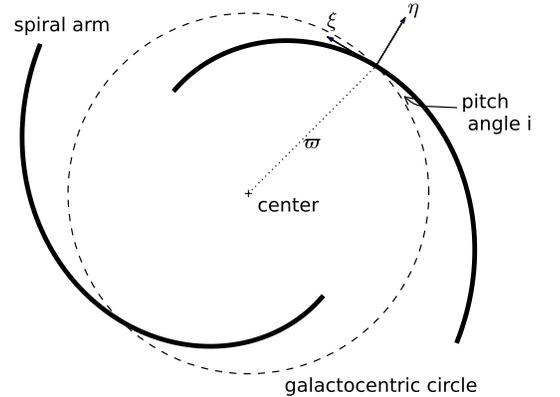}
\caption{Spiral coordinates $(\eta,\xi)$ are defined in the direction parallel and perpendicular to the spiral arm, locally at a galactocentric radius $\varpi$.}\label{spiral_coor}
\end{center}
\end{figure}

The two coordinate systems are related by the following metric:
\begin{equation}
ds^2=d\varpi^2+\varpi^2d\varphi^2=\frac{\varpi^2\sin^2 i}{m^2}(d\eta^2 +d\xi^2),
\end{equation}
which corresponds to a local rotation through an angle $i$ and rescaling of lengths by a common factor of $\varpi \sin{i}/m$.

\begin{figure}[!ht]
\begin{center}
\includegraphics[scale=0.45]{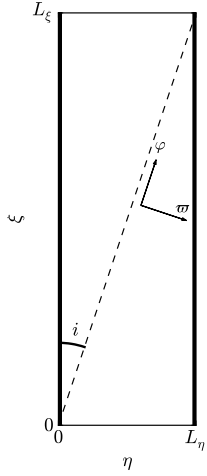}
\caption{A diagram showing the local rectangular box for the spiral coordinates $(\eta,\xi)$. The spiral arms are indicated by the bold vertical lines on two sides. The perpendicular distance between two arms is given by $L_{arm}\equiv L_\eta = 2\pi\varpi\sin{i}/m$, and the dash diagonal is the line constant galactocentric radius. Since the solution is doubly-periodic in both the $\eta$ and $\xi$ directions, coordinates $(0,0)$ and $(L_\eta, L_\xi)$ represent the same location, and $\tilde{L} \equiv \cot{i} =L_\xi / L_\eta $.}
\label{spiral_coor2}
\end{center}
\end{figure}

\subsection{Non-dimensionalization}
After the transformation into the spiral coordinate $(\eta,\xi)$, the equations of motion become
\begin{align}
\nonumber
\lefteqn{\frac{\varpi\sin{i}}{m}\left(\pd{u_{\eta 1}}{t}-2\Omega u_{\xi 1}\right)
+(u_{\eta 0}+u_{\eta 1})\pd{u_{\eta 1}}{\eta}}\\
+&(u_{\xi 0}+u_{\xi 1})\pd{u_{\eta1}}{\xi}=-\pd{}{\eta}(\mathcal{H}_{\rm g1}+\mathcal{H}_{\rm m1}+\mathcal{U}),
\end{align}
\begin{align}
\nonumber
\lefteqn{\frac{\varpi\sin{i}}{m}\left(\pd{u_{\xi 1}}{t}+\frac{\kappa^2}{2\Omega} u_{\eta 1}\right)
+(u_{\eta 0}+u_{\eta 1})\pd{u_{\xi 1}}{\eta}}\\\nonumber
+&(u_{\xi 0}+u_{\xi 1})\pd{u_{\xi 1}}{\xi}\\
=&-\pd{}{\xi}(\mathcal{H}_{\rm g1}+\mathcal{U})+\left(\frac{u_{\eta0}+u_{\eta1}}{u_{\xi 0}}\right)\pd{\mathcal{H}_{\rm m1}}{\eta}.
\end{align}
The corresponding continuity equation reads
\begin{align}
\nonumber
\frac{\varpi\sin{i}}{m}\pd{\Sigma_1}{t}+&\pd{}{\eta}[(\Sigma_0+\Sigma_1)(u_{\eta0}+u_{\eta1})]\\
+&\pd{}{\xi}[(\Sigma_0+\Sigma_1)(u_{\xi0}+u_{\xi1})]=0.
\end{align}
To write the equations in dimensionless form, we start by picking relevant velocity scales for the problem:
\begin{equation}
u\equiv u_{\eta 1}/(2UV)^{1/2}\quad\mathrm{and}\quad v\equiv u_{\xi 1}/V,
\end{equation}
where we have followed SMR by defining
\begin{equation}
U\equiv\frac{\varpi\Omega\sin{i}}{m}\quad\mathrm{and}\quad V\equiv\frac{\varpi\kappa^2\sin{i}}{2\Omega m},
\end{equation}
such that the Coriolis terms become $v$ and $-u$ for the $\eta$- and $\xi$-momentum equations, respectively. Similarly, we define $x_{\rm t0}^2\equiv v_{\rm t0}^2/2UV$ and $x_{\rm A0}^2\equiv v_{\rm A0}^2/2UV$ for the square of turbulent and Alfv\'en's speeds. For the record, we can rewrite the enthalpies into the dimensionless form:
\begin{align}
h_{\rm g} &\equiv \mathcal{H}_{\rm g 1}/(2UV)=-x_{\rm t0}/(1+\sigma), \\
h_{\rm m} &\equiv \mathcal{H}_{\rm m 1}/(2UV)=x_{\rm A0}(1+\sigma),
\end{align}
where $\sigma\equiv\Sigma_1/\Sigma_0$ is the relative gas surface density between the perturbed and axisymmetric states. We also rewrite the perturbed gravitational potential $\mathcal{U} \equiv \mathcal{V}_*+\mathcal{V}_{\rm g}$ as
\begin{equation}\label{totalpotential}
\mathcal{U}=\left(\frac{\sin{i}}{m}\right)[-(\varpi^2\Omega^2)F\cos{\eta}+(2\pi\varpi G\Sigma_0)\phi],
\end{equation}
where $\phi$ is the perturbed self-gravitational potential of the gas in units of $(\varpi \sin{i}/m) 2\pi G\Sigma_0$. Finally, we measure time in units of inverse epicyclic frequency:
\begin{equation}\label{tau}
d\tau \equiv \kappa dt.
\end{equation}

We introduce now the following additional dimensionless parameters:
\begin{align}
\nu &\equiv -u_{\eta 0}/(2UV)^{1/2}=m(\Omega_p-\Omega)/\kappa, \\
f &\equiv \left(\frac{\Omega}{\kappa}\right)^2\left(\frac{mF}{\sin{i}}\right), \\
\alpha &\equiv \frac{(\varpi \sin{i}/m) 2\pi G\Sigma_0}{2UV}=\frac{2\pi m G\Sigma_0}{\varpi \kappa^2 \sin{i}},
\end{align}
where $f$ is the afore-mentioned true dimensionless measure of the nonlinearity of the stellar forcing, and $\alpha$ is a similar dimensionless measure of the strength of the self-gravity of the gas.  Although $2\pi G \Sigma_0$ may be regarded as a small correction to the axisymmetric gravitational field of the galaxy, $\varpi\Omega^2$, nonlinear compressions behind galactic shocks make gas self-gravity a fierce  contractional competitor to the vortical spinup represented by $\varpi\kappa^2 \sin i$ when $\alpha$ is an order unity parameter.  

The continuity and momentum equations now take the dimensionless form:
\begin{align}\nonumber
&\pd{\sigma}{\tau}+\pd{}{\eta}\left[(1+\sigma)(-\nu+u)\right]\\
+&\pd{}{\xi}\left[(1+\sigma)\left(-\frac{\nu}{\tan i}+\frac{\kappa}{2\Omega}v\right)\right] = 0,\\
\nonumber
&\pd{u}{\tau} + (-\nu+u)\pd{u}{\eta} +\left(-\frac{\nu}{\tan i}+\frac{\kappa}{2\Omega}v\right)\pd{u}{\xi}\\
=& v-\frac{x_{\rm t0}}{(1+\sigma)^2}\pd{\sigma}{\eta}+f_\eta- \alpha \pd{\phi}{\eta} - f\sin{\eta},\\
\nonumber
&\pd{v}{\tau} + (-\nu+u)\pd{v}{\eta}+ \left(-\frac{\nu}{\tan i}+\frac{\kappa}{2\Omega}v\right)\pd{v}{\xi}\\
=&-u + \frac{2\Omega}{\kappa}\left[-\frac{x_{\rm t0}}{(1+\sigma)^2}\pd{\sigma}{\xi}-\alpha\pd{\phi}{\xi}\right]+f_\xi,
\end{align}
where $f_\eta$ and $f_\xi$ are the components of the dimensionless Lorentz force per unit mass in the directions perpendicular and along the spiral arm, respectively:
\begin{equation}
f_\varpi = {\mathcal{F}_\varpi\over (2UV)^{1/2}}, \qquad f_\varphi = {\mathcal{F}_\varphi\over V}.
\end{equation}
For the closure of the equations, we also need the solution of the Poisson's equation for the self-gravity of the gas and the equation of field freezing for the magnetic field.

\section{ONE-DIMENSIONAL SPIRAL SHOCK}
\label{spiralshock}
In this section we revisit the steady, 1-D, TASS solution, adding in the consideration of the effects of magnetic field (see also Roberts \& Yuan 1970) and self-gravity of the gas \citep[see also][]{1986ApJ...309..496L}. The TASS state, denoted by hats, gives the background flow of the feathering problem. All hatted quantities depend only on $\eta$, in the form $\hat{q}=\hat{q}(\eta)$. 

The Lorentz force per unit mass now reads,
\begin{equation}\label{lf1}
\hat{f}_\eta = -x_{\rm A0} \od{\shat}{\eta},
\end{equation}
and,
\begin{equation}\label{lf2}
\hat{f}_\xi = \frac{2\Omega}{\kappa}x_{\rm A0}\left(\frac{\tan{i}}{1+\shat}\right)\od{\shat}{\eta}.
\end{equation}
Note that the Lorentz acceleration in two directions differ in scale by an extra factor of $2\Omega/\kappa$ $(=\sqrt{2UV} / V)$ because of the difference in defining the dimensionless $u$ and $v$. Nevertheless, $f_\xi$ is smaller than $f_\eta$ by the factor $\tan i$, and it can be dropped asymptotically in the dynamical equation for $u_\xi$. By also dropping the derivatives in $\tau$ and $\xi$, 
the governing equations for velocity in $\eta$ and $\xi$ directions now read as follows:
\begin{equation}\label{c1}
	\od{\uhat}{\eta}=(-\nu+\uhat)\frac{\vhat-\alpha d\phat/ d\eta-f\sin{\eta} }{(-\nu+\uhat)^2-\xhat},
\end{equation}
\begin{equation}\label{c2}
	\od{\vhat}{\eta}=\frac{\uhat}{\nu-\uhat},
\end{equation}
where $\xhat \equiv x_{\rm t0}(1+\shat)^{-1} + x_{\rm A0} (1+\shat)$ is the square of the effective signal speed, and the parameters $f$ and $\alpha$ measure the relative strength of the stellar and gaseous perturbation gravitational fields. We make use of the dimensionless mass flux as a conserved quantity along the flow by integrating the continuity equation and putting it into the form:
\begin{equation}\label{c3}
	(1+\shat)(-\nu+\uhat)=-\nu.
\end{equation}

The self-gravity term can be obtained from a given surface density $\shat(\eta)$ by solving the Poisson equation under the WKBJ approximation. The derivation is standard and given in Appendix \ref{appendixA}.  The solution can be expressed in Fourier series form once $\shat(\eta)$ has been found by integrating the ODEs (\ref{c1} and \ref{c2} for $\uhat$ and $\vhat$, coupled with equation \ref{c3} for $\shat$):
\begin{equation}\label{sg1}
\shat = C_0 + \displaystyle\sum^{\infty}_{n=1} \left[ C_n \cos{(n \eta)} + S_n \sin{(n \eta)} \right], \\
\end{equation}
and
\begin{equation}\label{sg2}
\od{\phat}{\eta}= \displaystyle\sum^{\infty}_{n=1} \left[-S_n \cos{(n \eta)} + C_n \sin{(n \eta)} \right],
\end{equation}
where $C_n$ and $S_n$ are the $n$-th Fourier components for even and odd solutions, respectively. On the other hand, the integration of equation (\ref{c1}) requires knowledge of $\phat$, so iteration (with a relaxation parameter) is required to find numerically a completely self-consistent solution.   Apart from the added iteration and convergence steps for the self-gravity when $\alpha$ is nonzero, and a different expression for $\hat{x}$, the equations (\ref{c1},\ref{c2}) have the same form as the set studied by SMR, and they can be solved by using the same shooting method with the matching of upstream and downstream flows satisfying the shock jump conditions discussed below. 

\subsection{Magnetosonic Point and Shock Jump Conditions}
The spiral shock solution is periodic in the $\eta$ direction in the sense that when the flow passes through the shock front, it will accelerate from the submagnetosonic speed to supermagnetosonic speed, and eventually reach another shock at the next spiral arm. Thus, the region between two consecutive shocks is transmagnetosonic and has a magnetosonic point location ($\eta = \eta_{\rm mp}$), where the speed of the flow equals the local speed of magnetosound. Solutions with multiple magnetosonic points and shocks are also possible, but their study is beyond the scope of this paper (see SMR and \citealt{2003ApJ...596..220C} for discussions of the role of ultraharmonic resonances for producing spiral branches and their possible relationship to flocculence when overlapping resonance leads to chaotic nonlinear behavior). The magnetosonic point is located where the following condition is satisfied, 
\begin{equation}
	(-\nu+\uhat)^2-\xhat = 0,
\end{equation}
which is also an apparent singular point of the equation (\ref{c1}). By substituting equation ($\ref{c3}$) and the equation of state, we get
\begin{equation}\label{eqn_cubic}
	(-\nu+\uhat)^3+\frac{x_{\rm t0}}{\nu}(-\nu+\uhat)^2+ x_{\rm A0}\nu = 0,
\end{equation}
which is a cubic equation that gives only one positive value of $(-\nu+\uhat)=(-\nu+\uhat_{\rm mp})$ algebraically if $\nu$ is negative. A smooth solution across the magnetosonic point can be found by requiring both the numerator and denominator to be zero in the equation (\ref{c1}). Therefore, the derivatives of $\uhat$ and $\vhat$ at the magnetosonic point can be evaluated as:
\begin{equation}
	\od{\uhat}{\eta}\bigg|_{\rm mp} = \frac{ [ -\uhat_{\rm mp}/\hat{y}_{\rm mp}-\alpha \phat^{\prime\prime}|_{\rm mp}-f\cos{\eta_{\rm mp}} ]^{1/2} }{ \left[ 2 + x_{\rm t0}\nu^{-1}/\hat{y}_{\rm mp} - x_{\rm A0}\nu/\hat{y}_{\rm mp}^3 \right]^{1/2} },
\end{equation}
\begin{equation}
	\od{\vhat}{\eta}\bigg|_{\rm mp} = \frac{\uhat_{\rm mp}}{\nu-\uhat_{\rm mp}},
\end{equation}
where $\phat^{\prime\prime}$ is the second $\eta$-derivative of $\phat$, and we define $\hat{y}_{\rm mp} \equiv -\nu+\uhat_{\rm mp}$. Note that the value of the derivatives can be evaluated by solving $\uhat_{\rm mp}$ in advance from equation (\ref{eqn_cubic}) with the background parameters given. In our implementation of the shooting method, we start the integration from the neighboring points of the magnetosonic transition to the shock front in both supermagnetosonic and submagnetosonic directions separately. The ``initial'' values of $\uhat$ and $\vhat$ at these points are given by the following Taylor's series:
\begin{equation}
\begin{split}
	\uhat &= \uhat_{\rm mp} + \od{\uhat}{\eta}\bigg|_{\rm mp} (\eta-\eta_{\rm mp}) + \cdots \\
	\vhat &= \alpha\od{\phat}{\eta}\bigg|_{\rm mp} +f\sin{\eta_{\rm mp}} + \od{\vhat}{\eta}\bigg|_{\rm mp} (\eta-\eta_{\rm mp}) + \cdots.
\end{split}
\end{equation}
The derivatives of self-gravitational potential $\phat $ are obtained from the solution of previous step. Since $d\phat/d\eta$ is generally a smooth and continuous function, the value of $\phat^{\prime\prime}\vert_{\rm mp}$ may be expressed in a finite difference form with little numerical error.

\subsection{Matching Conditions}

The physical problem is constrained by the fact that the downstream and upstream flows for a periodic solution must match the values that allow a shock jump conditions to connect the supermagnetosonic and submagnetosonic collision. These jump conditions are obtained by requiring the sum of gas (turbulent) and magnetic pressures and momentum fluxes to be continuous across the shock:
\begin{equation}\label{eqn:rh1}\nonumber
\left[(1+\shat)(-\nu+\uhat)\uhat + x_{\rm t0} \ln(1+\shat) + \frac{x_{\rm A0}}{2}(1+\shat)^2\right]^2_1
\end{equation}
and,
\begin{equation}\label{eqn:rh2}\nonumber
\left[(1+\shat)(-\nu+\uhat)\vhat\right]^2_1
\end{equation}
to vanish separately. We can identify the constant mass flux, $(1+\shat)(-\nu+\uhat)=-\nu$ from the continuity equation above. Thus, we may obtain the corresponding post-shock (or pre-shock) values of $\uhat$ and $\vhat$ for a given pair of values on the other side of the shock. In practice, we calculate the corresponding post-shock (submagnetosonic) values by using the pre-shock (supermagnetosonic) values, as if they were to satisfy the shock jump conditions. We postpone the discussion of the numerical results until \S \ref{results}.

\section{FEATHERING ANALYSIS}
\label{featpert}
In the context of the feathering phenomenon, we need to consider variations in 2-D and time. Because the reference TASS state is independent of $\xi$ and $\tau$, we may describe, in a linearized treatment, the additional feathering variations as oscillatory disturbances in $\xi$ and $\tau$. Such a treatment constitutes a standard linear-stability analysis and should be contrasted with prior treatments that supposed feathering to be a shearing, time-dependent, phenomenon that is imposed by the mathematics of a transformation that is useful only near corotation. In the current paper, we deliberately stay away from corotation.  A complication that does appear is the oscillations introduced by a wiggling shock front, which leads to perturbed jump conditions that further affects the downstream flow. In any case, instead of solving a set of PDEs as in the numerical experiments (which do not need linearization), we obtain a set of ODEs for the feathering perturbation.  Imposing double-periodicity, for given $m$ (= 2 in the usual TASS picture) and $l$ (the number of feathers strung out along the arms in the $\xi$-direction per spiral arm box), the (complex) oscillation frequency $\omega_R+i\omega_I$ in time becomes an eigenvalue of the overall problem. 

\subsection{Perturbational Equations}

As the feathering perturbations are time-dependent and vary along both $\eta$ and $\xi$ directions spatially, we define the variables as follows:
\begin{equation}
\begin{split}
	u &= \uhat (\eta) + \tilde{u} (\eta,\xi,t), \\
	v &= \vhat (\eta) + \tilde{v} (\eta,\xi,t), \\
	\sigma &= \shat (\eta) + \tilde{\sigma} (\eta,\xi,t), \\
	\phi &= \phat (\eta) + \tilde{\phi} (\eta,\xi,t), \\
\end{split}
\end{equation}
where the hat states are the background TASS flow, and the tilde states are perturbations assumed to be small compared to the background. The perturbational magnetic field is time-dependent, and we no longer assume its direction is parallel to the flow as was assumed for the TASS background. We will derive the perturbation induction equation in Appendix \ref{appendixB}), where we show it corresponds simply to the conservation relation for the magnetic flux function $A$ ($z$-component of the vector potential for the magnetic field). For here, we simply record the resulting linearized perturbation fluid equations for the tilde quantities:
%
\begin{align}
\label{f21}\nonumber
\pd{\stilde}{\tau} &+ (-\nu+\uhat)\pd{\stilde}{\eta} + \utilde \od{\shat}{\eta}+\hat{v}_{\rm T} \pd{\stilde}{\xi} 
\\
=&-(1+\shat)\pd{\utilde}{\eta}-\od{\uhat}{\eta}\stilde-(1+\shat)\frac{\kappa}{2\Omega}\pd{\vtilde}{\xi},
\end{align}
\begin{align}
\label{f22}
&\pd{\utilde}{\tau} + (-\nu+\uhat)\pd{\utilde}{\eta} + \utilde \od{\uhat}{\eta}+\hat{v}_{\rm T} \pd{\utilde}{\xi}
\\\nonumber
=& \vtilde-\left[\frac{x_{\rm t0}}{(1+\shat)^2}\right]\pd{\stilde}{\eta}+\od{\shat}{\eta}\frac{2x_{\rm t0}}{(1+\shat)^3}\stilde-\alpha \pd{\ptilde}{\eta}+\tilde{f}_\eta,
\end{align}
\begin{align} 
\label{f23}\nonumber
&\pd{\vtilde}{\tau} + (-\nu+\uhat)\pd{\vtilde}{\eta} + \utilde \od{\vhat}{\eta}+\hat{v}_{\rm T} \pd{\vtilde}{\xi} 
\\
=&-\utilde-\frac{2\Omega}{\kappa}\left[\frac{x_{\rm t0}}{(1+\shat)^2}\pd{\stilde}{\xi}+ \alpha \pd{\ptilde}{\xi}\right]+\tilde{f}_\xi,
\end{align}
where $\hat{v}_{\rm T}$ is the total $\xi$-component of the background fluid velocity (i.e., axisymmetric plus TASS) in the frame that corotates with the stellar spiral. The perturbed Lorentz acceleration, $\tilde{f}_\eta$ and $\tilde{f}_\xi$ are given by
\begin{align}
\tilde{f}_\eta = x_{\rm A0}\left(\frac{\partial^2}{\partial\eta^2}+\frac{\partial^2}{\partial\xi^2} \right) \tilde{A}_1 +x_{\rm A0}\frac{\shat'}{1+\shat}\pd{\tilde{A}_1}{\eta},
\end{align}
and,
\begin{align}
\nonumber
\tilde{f}_\xi =& -\frac{2\Omega}{\kappa}x_{\rm A0}\left(\frac{\tan{i}}{1+\shat}\right)\left(\frac{\partial^2}{\partial\eta^2}+\frac{\partial^2}{\partial\xi^2} \right)\tilde{A}_1\\
+&\frac{2\Omega}{\kappa}x_{\rm A0}\frac{\shat^\prime}{1+\shat}\frac{\partial{\tilde{A}_1}}{\partial\xi} ,
\end{align}
respectively.  In the above, $\tilde{A}_1$ is the dimensionless $z$-component of the perturbational magnetic potential. The perturbational evolutionary equation for it reads (see Appendix B):
\begin{align}\label{f24}
\nonumber
&\pd{\tilde{A}_1}{\tau} + (-\nu+\uhat) \pd{\tilde{A}_1}{\eta} + \hat{v}_{\rm T} \pd{\tilde{A}_1}{\xi}
\\
=& (1+\shat)\utilde - \left(\frac{\kappa}{2\Omega}\tan{i}\right)\vtilde.
\end{align}

\subsection{Perturbed shock jump conditions}
\begin{figure}
\begin{center}
\includegraphics[scale=0.4]{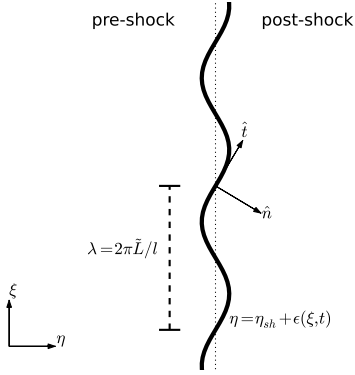}
\caption{Corrugation of the shock front. The perturbation, $\epsilon$, along the shock front is characterized by the wavelength $\lambda=2\pi\tilde{L}/l$ and is assumed to have small amplitude. The vertical dotted line is equilibrium shock front at $\eta =\eta_{sh}$. The unit normal and unit tangent are denoted by $\hat{n}$ and $\hat{t}$, respectively.}\label{FIG_CORR}
\end{center}
\end{figure}
The perturbed shock jump conditions can be obtained by linearizing the jump conditions in the frame of perturbed shock front. The shock front is displaced and no longer parallel to the spiral arm. In Figure \ref{FIG_CORR}, we show the configuration of the perturbation. Inside corotation radius, the flow is entering the shock from the left in the frame of the shock. Hence, we must obtain the normal direction of the shock and the shock velocity in the current frame. The position of the perturbed shock is now given by,
\begin{equation}
	\eta = \eta_{sh} + \epsilon(\xi,t),
\end{equation}
where $\eta_{sh}$ is the unperturbed position and $\epsilon$ is a small number. We define,
\begin{equation}
	s = \eta - \eta_{sh} - \epsilon(\xi,t),
\end{equation}
to be the displacement from the moving shock front. Thus, the locus of the shock front is $s=0$ and the unit normal $\hat{\mathbf{n}}$ is given by,
\begin{equation}\label{unitnormal}
\begin{split}
	\hat{\mathbf{n}} &= \frac{1}{|\nabla s|}\left[ \left(\pd{s}{\eta}\right) \hat{\mathbf{e}}_\eta + \left(\pd{s}{\xi}\right) \hat{\mathbf{e}}_\xi \right] \\
	&= \frac{1}{|\nabla s|}\left[ \hat{\mathbf{e}}_\eta - \left(\pd{\epsilon}{\xi}\right) \hat{\mathbf{e}}_\xi \right],\\
\end{split}
\end{equation}
where $|\nabla s| = [ 1 + (\partial{\epsilon}/\partial{\xi})^2 ]^{1/2} \simeq 1 + O(\epsilon^2)$ is the magnitude of the gradient normal. The velocity of the shock, $\mathbf{D}$, which is normal to the shock front, can be found by considering a normal displacement $\Delta r$ of the locus at time $\Delta t$, i.e., $|\nabla s| \Delta r + \Delta t (\partial s/\partial t) = 0$, and thus, $\mathbf{D} = \hat{\mathbf{n}} \Delta r/ \Delta t = (\partial \epsilon/ \partial t) \hat{\mathbf{n}}$ in the first order of $\epsilon$.

There are five shock jump conditions in the problem \citep[see][eqs 25.16-18,20,21]{1992pavi.book.....S}. The linearized perturbation in the moving shock frame reads,
\begin{align}
&\left[\rho\delta u_\bot+u_\bot\delta\rho \right]^2_1 = 0,
\label{pbc01}\\
&\left[u^2_\bot\delta\rho+2\rho u_\bot\delta u_\bot + \left(\frac{\partial P}{\partial \rho}\right)\delta\rho+\frac{B_\|}{4\pi}\delta B_\|\right]^2_1 = 0,
\label{pbc02}\\
\nonumber
&[u_\bot u_\| \delta\rho +\rho u_\bot \delta u_\| +\rho u_\| \delta u_\bot
\\
&-\frac{B_\bot}{4\pi}\delta B_\|-\frac{B_\|}{4\pi}\delta B_\bot]^2_1= 0,
\label{pbc03}\\
&[\delta B_\bot]^2_1 = 0,
\label{pbc04}\\
&\left[B_\bot \delta u_\| - B_\| \delta u_\bot + u_\| \delta B_\bot - u_\bot \delta B_\|\right]^2_1 = 0,
\label{pbc05}
\end{align}
where the variables with and without $\delta$ are in first and zeroth order of $\epsilon$, respectively. The jump conditions are given in terms of variables parallel and perpendicular to the shock front locus, $s=0$. Since the perturbation on the shock front is in first order of $\epsilon$, the non-$\delta$ variables are simply their background counterpart. For the variables that are first order in $\epsilon$, we include both Taylor's expansion at the perturbed shock front and the geometrical projection (i.e., in Lagrangian sense). Thus, by making use of the unit normal in equation (\ref{unitnormal}), and the corresponding unit tangent, we express the dimensional gas surface density, flow velocities and magnetic fields into the following,
\begin{align}
\rho &\simeq \hat{\rho}+\tilde{\rho}+\epsilon d\hat{\rho}/d\eta,
\label{app1}\\
u_\bot &\simeq \hat{u}_\eta +\tilde{u}_\eta+\epsilon d\hat{u}_\eta/d \eta-D_\eta-\hat{u}_\xi \partial_\xi \epsilon ,
\label{app2}\\
u_\| &\simeq \hat{u}_\xi+\tilde{u}_\xi+\epsilon d\hat{u}_\xi/d \eta+\hat{u}_\eta \partial_\xi \epsilon,
\label{app3}\\
B_\bot &\simeq  \hat{B}_\eta+\tilde{B}_\eta+\epsilon d\hat{B}_\eta /d\eta-\hat{B}_\xi \partial_\xi \epsilon,
\label{app4}\\
B_\| &\simeq  \hat{B}_\xi+\tilde{B}_\xi+\epsilon d\hat{B}_\xi /d\eta + \hat{B}_\eta \partial_\xi \epsilon,
\label{app5}
\end{align}
where we evaluate the variables at $\eta=\eta_{sh}$, and we define $D_\eta$ and $D_\xi$ as the $\eta$ and $\xi$ components of the shock velocity, respectively. Note that $D_\eta=(\partial{\epsilon}/\partial{t})/|\nabla s|^2\simeq(\partial{\epsilon}/\partial{t})$, and we exclude $D_\xi = O(\epsilon^2)$. To make the equations dimensionless, we measure the surface density, velocities and magnetic fields in term of $\Sigma_0$, $\sqrt{2UV}$ and $B_{\xi0}$, respectively. We now write the boundary conditions (\ref{pbc01} - \ref{pbc05}) a more compact form,
\begin{equation}\label{bc1}
\textbf{Q}(1)\textbf{V}(1) = \textbf{Q}(2) \textbf{V}(2),
\end{equation}
where $\mathbf{Q}=\mathbf{Q}(\eta)$ is a $5\times 5$ matrix is given by the coefficients in the equations (\ref{app1}-\ref{app5}) in terms of the background variables and $\mathbf{V}= [\delta\sigma, \delta u, (\kappa/ 2\Omega) \delta v, \delta B_\bot/B_{\xi0}, \delta B_\|/B_{\xi0}]^T$ is a column vector of the $\delta$ variables.

\subsection{Stability analysis}
\label{SA}
As the governing equations for the perturbed variables do not have explicit dependence in $\tau$ and $\xi$,  we can simplify the equations by assuming that the tilde variables have $e^{i\omega \tau - il\xi/\tilde{L}}$ dependences.  The perturbed shock front must have the same sinusoidal dependence in time and space (see Figure \ref{FIG_CORR}).  The instability condition follows when $\mathrm{Im} (\omega)=\omega_{\rm I} < 0$. To treat the perturbational self-gravity, we use a simplified solution to the Poisson equation obtained in Appendix \ref{appendixA}:
\begin{equation}
	\ptwl(\eta) = -\frac{\tilde{L}}{|l|}\stwl(\eta),
\end{equation}
appropriate to the assumption (motivated by the observations) that the important feathering corresponds to large $l$ (many individual feathers in the $\xi$ direction for the equivalent spiral box where we only have 1 TASS arm becoming another in the $\eta$ direction).  

After rearranging the terms, we get
\begin{align}\label{s01}\nonumber
&(-\nu+\uhat)\od{\stwl}{\eta} + (1+\shat)\od{\utwl}{\eta} \\
=& (-\uhat'-i\omega_{\rm T})\stwl- \shat'\utwl +\frac{il}{\tilde{L}}\frac{\kappa}{2\Omega}(1+\shat)\vtwl,
\end{align}
\begin{align}\label{s02}\nonumber
&\left[\frac{x_{\rm t0}}{(1+\shat)^2}-\alpha\frac{\tilde{L}}{|l|}\right]\od{\stwl}{\eta}\\\nonumber
&+ (-\nu+\uhat)\od{\utwl}{\eta}-\frac{x_{\rm A0}\shat'}{1+\shat} \od{\tilde{A}_{1\omega,l}}{\eta}-x_{\rm A0}\od{\tilde{A}^{\prime}_{1\omega,l}}{\eta} \\
=& \frac{2x_{\rm t0}\shat'}{(1+\shat)^3}\stwl + (-\uhat'-i\omega_{\rm T})\utwl+\vtwl \\\nonumber
&-x_{\rm A0}\left(\frac{l}{\tilde{L}}\right)^2 \tilde{A}_{1\omega,l},
\end{align}
\begin{align}\label{s03}\nonumber
&(-\nu+\uhat)\od{\vtwl}{\eta} + \frac{2\Omega}{\kappa}x_{\rm A0}\left(\frac{\tan{i}}{1+\shat}\right)\od{\tilde{A}^{\prime}_{1\omega,l}}{\eta}  \\\nonumber 
=&-\left(-\frac{il}{\tilde{L}}\right)\left(\frac{2\Omega}{\kappa}\right)\left[ \frac{x_{\rm t0}}{(1+\shat)^2}-\alpha\frac{\tilde{L}}{|l|}\right]\stwl\\\nonumber
&-(1+\shat)\utwl -i\omega_{\rm T}\vtwl \nonumber \\
&+ \frac{2\Omega}{\kappa}x_{\rm A0}\left[\frac{\tan{i}}{1+\shat}\left(\frac{l}{\tilde{L}}\right)^2-\frac{\shat^{\prime}}{1+\shat}\left(\frac{il}{\tilde{L}}\right)\right]\tilde{A}_{1\omega,l},
\end{align}
where we define $\omega_{\rm T} \equiv \omega - (l/\tilde{L})\hat{v}_{\rm T} $ to be the dimensionless Doppler-shifted frequency in the moving frame of the background flow along the spiral arm. The transformed induction equation reads,
\begin{align}\label{s04}
&(-\nu+\uhat)\tilde{A}^{\prime}_{1\omega,l}\\\nonumber
=&-i\omega_{\rm T}\tilde{A}_{1\omega,l} + (1+\shat)\utwl-\left(\frac{\kappa}{2\Omega}\tan{i}\right)\vtwl.
\end{align}
Similarly, by taking the Fourier transform and matching the first order terms in the perturbational jump conditions (\ref{app1}-\ref{app5}), we can express the $\delta$ terms in the in terms of the tilde (Eulerian) variables:
\begin{align}\nonumber
\delta \sigma &= \stilde +\epsilon d\shat /d\eta, \\\nonumber
\delta u &= \utilde + \epsilon d\uhat /d\eta-i\omega_{\rm T}\epsilon, \\
\delta v &= \vtilde + \epsilon d\vhat/d\eta-\frac{2\Omega}{\kappa}\frac{il}{\tilde{L}}\uhat\epsilon, \\\nonumber
\delta B_\bot / B_{\xi0} &= -\frac{il}{\tilde{L}}\tilde{A}_1+\frac{il}{\tilde{L}}(1+\shat)\epsilon,\\\nonumber
\delta B_\| / B_{\xi0} &= -\tilde{A}^{\prime}_1+\epsilon d\shat/d\eta,
\end{align}
where we drop the subscripts $(\omega,l)$ for clarity, and we have used $d\hat{B}_\eta/d\eta \propto d(\Sigma u_\eta)/d\eta=0$,
\begin{equation}\nonumber
\frac{\epsilon}{B_{\xi0}}\od{\hat{B}_\xi}{\eta} = \frac{\epsilon}{\Sigma_0}\od{\Sigma}{\eta} \simeq \epsilon \od{\shat}{\eta},
\end{equation}
\begin{equation}\nonumber
\frac{B^{(0)}_\|}{B_{\xi0}} \simeq \frac{\hat{B}_\xi}{B_{\xi0}} = (1+\shat) \quad\text{and}\quad \frac{B^{(0)}_\bot}{B_{\xi0}} \simeq \frac{\hat{B}_\eta}{B_{\xi0}} = \tan{i},
\end{equation}
and we neglect the term with $\partial_\xi \epsilon (\hat{B}_\eta/B_{\xi0}) = O(\epsilon \tan{i})$.

\subsection{Method of solution}
\label{mos}
We try to solve the \textit{four} perturbed equations (\ref{s01}-\ref{s04}) as a set of ODEs. Since the Lorentz force contains a second derivative of the scalar magnetic potential, it might seem that we require one more differential equation for its derivative, $d\tilde{A}_1/d\eta = \tilde{A}_1^\prime$. However, the induction equation (\ref{s04}) is in fact an algebraic equation for $\utilde$, $\vtilde$, $\tilde{A}_1$ and $\tilde{A}_1^\prime$, and does not involve $d\tilde{A}_1^\prime/d\eta$. The physical reason is that field freezing implies that the tilde magnetic potential must yield a magnetic field structure that corresponds to the TASS magnetic field stretched in time by the motion of the electrically conducting matter in the feathering instability (see Appendix B). In other words, the field freezing equation (\ref{s04}) must hold in space simultaneously with the other differential equations, and so, the system is a set of \textit{Differential Algebraic Equations} (DAEs), where $d\tilde{A}_1^\prime/d\eta$ is obtained by {\it differentiation} after we make the set self-consistent by treating $\omega$ not as an arbitrary constant, but as a (complex) eigenvalue of the time-dependent transformation of the TASS state to one that contains (exponentially growing) feathering perturbations. Fortunately, the problem so posed can be solved by following the procedure discussed below.

First, we should reduce the order of the perturbed equations such that it solves for the tilde variables $( \stilde, \utilde, \vtilde, \tilde{A}_1 )$ (and their first derivatives on the RHS) only. The solution of $\tilde{A}_1^\prime$ can be obtained algebraically from the induction equation once we have solved for other variables. The elimination of $\tilde{A}_1^{\prime\prime}$ may then be done by numerically differentiation of the induction equation, i.e.,
\begin{align}\label{s05}\nonumber
&-(1+\shat)\od{\utwl}{\eta} + \frac{\kappa}{2\Omega}\tan{i}\od{\vtwl}{\eta} + (-\nu+\uhat)\od{\tilde{A}^{\prime}_{1\omega,l}}{\eta} \\
=& \shat'\utwl+\left(\frac{il}{\tilde{L}}\right)\left(\frac{\kappa}{2\Omega}\right)(1+\shat)\tilde{A}_{1\omega,l}-(\uhat'+i\omega_{\rm T})\tilde{A}^{\prime}_{1\omega,l}.
\end{align}
Second, we should reduce the number of perturbational jump conditions to \textit{four}, as we have \textit{four} differential equations after the reduction above. In fact, the fifth jump condition, equation (\ref{pbc05}), is satisfied automatically by the ``algebraic" induction equation. In other words, if the complex frequency $\omega$ has the correct eigenvalue, we can impose both double-periodicity and match all the requisite jump conditions, with the equation (\ref{pbc05}) being consistent with the induction equation with which we use to calculate $\tilde{A}_1^\prime$. We can eliminate $\tilde{A}_1^\prime$ in the connection conditions by using the equation (\ref{s04}), which is valid for both sides of the shock separately. Therefore, the dimension of the coefficient matrix \textbf{Q} in equation (\ref{bc1}) is reduced to $4\times4$. 

After we ``separate" the induction equation from the differential equations in the above manner, we may solve the system as a \textit{Two-Point Boundary Value Problem with Eigenvalues}. Standard methods of attack exist in the literatures, e.g., \cite{ascher1995numerical}. One difficulty of using the publicly available numerical packages is that they do not treat the embedded jump conditions present in our system. To solve this problem, we can artificially modify our jump conditions in the following form:
\begin{equation}
\mathbf{Q(1)}\mathbf{V(1)}=\mathbf{C}_1 \quad \text{and} \quad \mathbf{Q(2)}\mathbf{V(2)}=\mathbf{C}_2,
\end{equation}
where the two vectors $\mathbf{C}_1$ and $\mathbf{C}_2$ are varied until they are equal to each other.  The procedure leads to the ability to use standard packages at the numerical expense of solving four more equations. There are also standard methods to solve a system with complex variables as in our problem, and we do not discuss the numerical issues further here.

\section{NUMERICAL RESULTS}
\label{results}
In this section we present results of the calculation with input parameters based on the \cite{1999ApJ...523..136S} rotation curve of the M81 Galaxy.
Table \ref{table1} lists the relevant numerical values used for the feathering calculation, and Table \ref{table2} gives the seven dimensionless parameters needed for solving the governing equations in that calculation.  The listed mean gas surface density $\Sigma_0$, gas turbulent velocity $v_{\rm t0}$, and interstellar magnetic field are all too high for a Sb spiral galaxy of luminosity class II like M81, but we adopt these extreme values to illustrate a point that will become apparent in our closing discussion.

In any case, the mean plasma $\beta$ is $\beta_0 \equiv x_{\rm t0}/x_{\rm A0}$ = 1 for the choice $x_{\rm t0} = x_{A0}$, and implies that turbulent stresses $\propto x_{\rm t0}/(1+\hat\sigma)$ dominate in the inter-arm region where $\hat\sigma < 0)$ while magnetic stresses $\propto x_{\rm A0} (1+\shat)$ dominate in the arm region where $\hat \sigma > 0$.  We define a mean Toomre's $Q$ parameter for the gas accounting for mean turbulent motions and magnetic field (with mean effective sound speed, $a_0 \propto \sqrt{x_0}$) as:
\begin{equation}
Q \equiv \frac{\kappa a_0}{\pi G\Sigma_0 } = \frac{2}{\alpha}(x_{\rm t0}+x_{\rm A0})^{1/2}.
\end{equation}
With $x_{\rm t0} = x_{A0} = 0.1$ and $\alpha = 0.35$, we have $Q = 2.55$, so the gas is stable on average to all axisymmetric self-gravitational perturbations. 

\begin{table}
\begin{center}
\caption{Typical galactic parameters for the feathering example of this paper}\label{table1}
\begin{threeparttable}
\begin{tabular}{cc}
\hline  Rotation curve & \cite{1999ApJ...523..136S} \\ 
Galactocentric radius \tnote{a} & 5.0 kpc \\
$L_{arm}$ & 3.8 kpc \\
Turbulent speed & 13.4 km/s \\
Alfv\'en's speed & 13.4 km/s \\
Magnetic field \tnote{b,c} & 14.8 $\mathrm{\mu G}$ \\
Mean gas surface density \tnote{c} & 38.6 $\mathrm{M_{\sun}/pc^2}$ \\
Inclination of stellar spiral arm \tnote{d} & $14^\circ $ \\
Pattern speed \tnote{e} & 23.4 km/s/kpc \\
\hline 
\end{tabular}
\begin{tablenotes}
	\item[a]{the distance from the modelling region to the galactic center}
	\item[b]{value of $B_{\varphi 0}$ for scale height $z_0=200pc$}
	\item[c]{for $\alpha=0.35$}
	\item[d]{adopted from \citet{2008MNRAS.387.1007K} }
	\item[e]{adopted from \citet{1998ApJS..115..203W} }
\end{tablenotes}
\end{threeparttable}
\end{center}
\end{table}

%
%
\begin{table}
\begin{center}
\caption{Typical set of local parameters ($\varpi = 5.0 {\rm kpc}$) for the feathering example of this paper. }\label{table2}
\begin{threeparttable}
\begin{tabular}{cc}
\hline  
$\Omega/\kappa$ & 0.666 \\ 
$\tan{i}$ & 0.249 \\
$F$ & 11.5$\%$ \\
$\alpha$ & 0.35 \\
$\nu$ & -0.666 \\
$x_{\rm A0}$ & 0.1 \\
$x_{\rm t0}$ & 0.1 \\
\hline 
\end{tabular}
\end{threeparttable}
\end{center}
\end{table}

\subsection{TASS profiles with self-gravity and magnetic field}

\begin{figure}[!ht]
\begin{center}
\includegraphics[scale=0.42,trim=20 0 0 0]{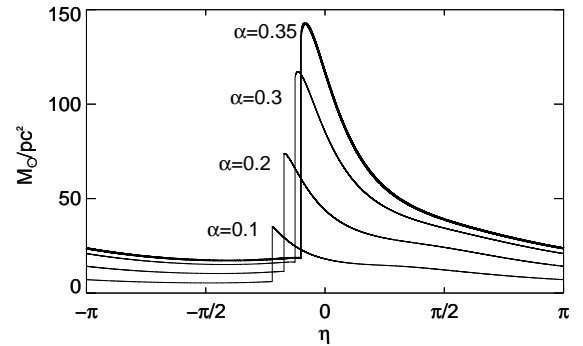}
\caption{Surface density of the gas for $\alpha$ from 0.1 to 0.35 (corresponding to 11 to 39 $\mathrm{M_\sun /pc^2}$ for the mean surface density) and $F=11.5\%$. The minimum stellar gravitational potential is located at $\eta=0$. The thick line ($\alpha=0.35$) is the background density profile used for the feathering perturbation example.}
\label{FIG_SHOCK}
\end{center}
\end{figure}

The dimensionless parameter $\alpha$ characterizes the gaseous self-gravity. In general, it is an order unity quantity for spiral galaxies of not too early a Hubble type. Figure \ref{FIG_SHOCK} shows the surface density in the perpendicular direction to the arm. The plot is similar to the \cite{Roberts1969a} calculation, except the surface density is no longer arbitrarily scalable when we include self-gravity. The shock strength increases with increasing $\alpha$ (with all other parameters held fixed as in Table 2) because the gaseous self-gravity deepens the minimum of the spiral gravitational potential.  More of the support for the total spiral gravitational potential coming from the gas also pulls the shock front downstream closer to the potential minimum.  Increases in $\alpha$ also rounds out the density peak. These effects were also seen in the 1-D numerical simulations of \cite{KO2002}.  

On the other hand, the full-width-at-half-maximum of the density profile is approximately constant. If the FWHM is taken as a representative value, the width of the arm with spiral galactic shocks is about $12\% $ of the distance between two arms, or about 480 pc in the current model.

\subsubsection*{Magnetic field}

\begin{figure}[!ht]
\begin{center}
\includegraphics[scale=0.42,trim=20 0 0 0]{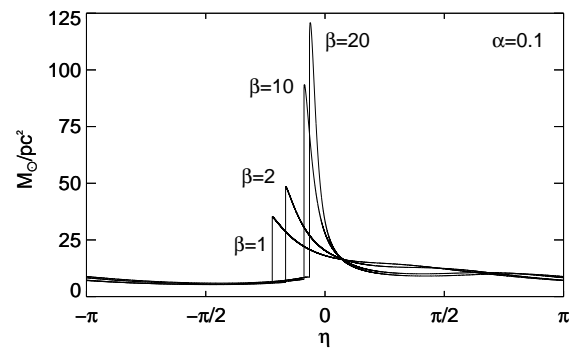}
\caption{Surface density of the gas for plasma $\beta$, $\beta_0 = 1.0, 2.0, 10$ and 20 (corresponding to $B_{\varphi0} =  7.9, 5.6, 2.5$ and 1.8 $\mu \rm G$, respectively). The mean surface gas density is set to 11 $\mathrm{M_\sun /pc^2}$ ($\alpha = 0.1$). }\label{FIG_BETA}
\end{center}
\end{figure}

Contrary to previous assertions \citep[cf.][]{2006MNRAS.367..873D}, the magnetic field plays an important role in spiral galactic shocks and the resultant feathering instabilities \citep[see, e.g.,][]{KO2002}.  Figure \ref{FIG_BETA} shows some different choices for the magnetization parameter $x_{A0} = x_{\rm t0}/\beta_0$, when the turbulent and self-gravity parameters are kept fixed at $x_{\rm t 0} = 0.1$ and $\alpha = 0.1$ with all other dimensionless parameters held at the values given in Table 2. In general, the increase in $x_{A0}$ (or decrease in $\beta_0$) suppresses the compression of gas in the postshock region, and lowers the shock strength. Conversely, the peak surface density rises very rapidly with increasing $\beta_0$ (i.e., decreasing magnetic field strength), and there is no steady solution possible for $\beta_0$ much larger than 20; i.e., the spiral arms would go into continued gravitational collapse with $\alpha = 0.1$ if the magnetic field is too weak. The magnetic field cannot be ignored either for the structure of the TASS density pattern or for the development of the feathering instability when self-gravity is important. The dependences of the background profile and the feathering perturbation on the dimensionless parameters of the problem will be investigated in Paper II.

\subsection{Feathering Perturbation}

\begin{figure}[!ht]
\begin{center}
\includegraphics*[scale=0.6,trim=100 0 0 0]{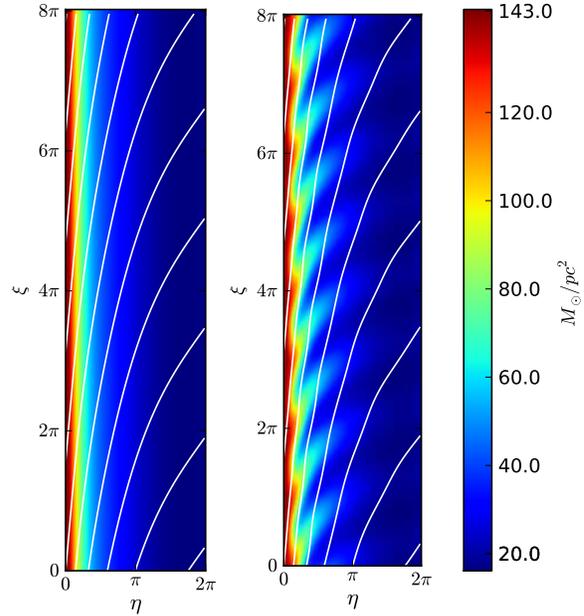}
\caption{Comparison between the background flow  with $\alpha=0.35$ (\textit{left}) and the flow with feathering perturbation  of the $l=8$ mode (\textit{right}). The color shows the surface density of the gas in linear scale, and the white lines are the magnetic field lines. Since $\tilde{L}=\cot{i}\simeq 4$ in the model, the periodicity of $\xi$ is approximately $8\pi$.}
\label{FIG_COMPARE}
\end{center}
\end{figure}

Taking the TASS 1-D solution as the background for the feathering phenomenon, we solve the perturbed equations for each $l$-mode and obtain the 2-D solution using inverse Fourier transforms. A larger survey of parameter space is undertaken in Paper II. Here, we just show a typical result for the total surface density and magnetic field lines in Figures \ref{FIG_COMPARE}.  To obtain sufficient contrast, we have arbitrarily scaled the linear perturbations so that they are no longer small compared to the background. Figure \ref{FIG_COMPARE} compares the flow solutions with and without the feathering perturbation for $l = 8$.  For better viewing of the post-shock region, the horizontal axis is not really $\eta$, but $\eta-\eta_{\rm sh}$.  For the background flow on the left, the magnetic field reaches peak compression behind the shock but become weaker for increasing $\eta$ as the expansional flow out of the spiral arm pulls apart the the frozen-in field lines.  With the development of the feathering perturbation with $l = 8$ on the right (i.e., 8 feathers in a distance, $L_\xi$), over-dense regions jut out from the spiral arm toward the interarm region downstream. In this particular case, the Doppler-shifted frequency is $\omega_{\rm T} = -0.113 -0.174 i$ in unit of $\kappa$. The non-zero real part of $\omega_{\rm T}$ implies that the pattern of feathers moves along the {\it outward} $\xi$ direction with the passage of time, a result also seen in the numerical calculations of the Ostriker group.  The negative value of the imaginary part of $\omega_{\rm T}$ implies that this mode is unstable and can be expected indeed to grow to nonlinear amplitudes with the passage of time. With $\kappa = 70.2$ km s$^{-1}$ kpc$^{-1}$, the $e$-folding growth time-scale $t_g$ is
\begin{equation}\label{growthtime}
t_g =-\frac{1}{\kappa \operatorname{Im}{\omega_T}}=\frac{1}{0.174 \kappa}\simeq 80 \times 10^6 \mathrm{yr}.
\end{equation}

\begin{figure}[!ht]
\begin{center}
\includegraphics*[scale=0.45,trim=20 0 0 0]{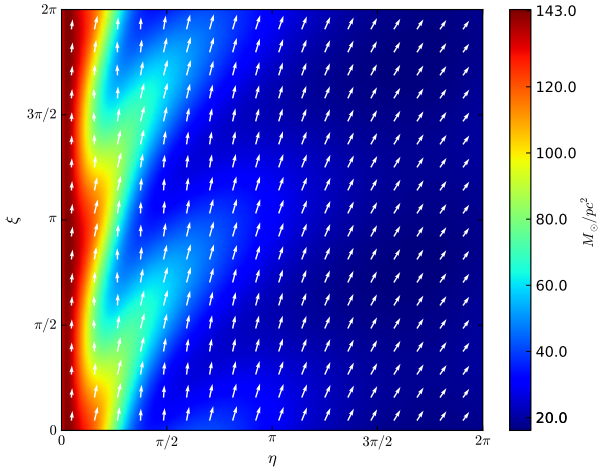}
\caption{Velocity field of the feathering perturbation $(l=8)$ with the background spiral shock.  Without feathering perturbation, the velocity is expected to have no variation along $\xi$.  }\label{FIG_VEL}
\end{center}
\end{figure}

In Figure \ref{FIG_VEL}, we show the velocity field (arrows) superimposed on the surface density profile (colored contours) of the feathering perturbation at a single instant of time.  The perturbed flow follows closely to the background spiral shock profile because the background circular motion is dominant over all other motions. Nevertheless, significant convergence toward density peaks and divergences from density troughs can still be found {\it along} the spiral arm, especially toward the beginning of the feathers. The behavior can be profitably compared to the zoom-in plot of \cite{2006ApJ...647..997S}.  It is tempting to speculate whether the velocity fluctuations in the nonlinear development of the instability can be mistaken for turbulent velocities in insufficiently angularly resolved images of spiral galaxies.

\section{DISCUSSIONS AND CONCLUSION}
\label{discussion}
It is illuminating to compare the numerical result (\ref{growthtime}) with a simple model of 1-D Jeans instability.  The Jeans instability cannot occur in 1-D equilibrium states if the compression occurs isothermally because the increase in  pressure forces keeps pace with the increase in self-gravitation \citep[cf.][]{1968nim..book....1S}, but if the (turbulent) equation of state is softer than isothermal, as it is for the adopted logatropic law of the current paper, then it is possible for the self-gravity of condensations parallel to the galactic shock to overwhelm the declining resistance of the turbulent motions. According to the estimate by \citet{2007ApJ...662L..75S}, the feathering instability is then basically the self-gravitational contraction of over-dense gas along post-shock magnetic field lines, which is almost aligned along the density-ridge of the TASS pattern.
%
%
%
%
A rough estimate of the contraction time is then
\begin{equation}
t_c \simeq \frac{\pi}{4} \left(\frac{G\Sigma_g W}{L^2}\right)^{-1/2},
\end{equation}
where $W$ is the half-width of the TASS spiral arm, which we shall take to equal the 0.48 kpc mentioned previously, and $L=L_{arm}(\tilde{L}/l)/4$ is the quarter-wavelength of the feathering instability (the center to edge distance of over-dense regions) and also equals, by coincidence, 0.48 kpc. If we replace $\Sigma_g$ with the mean surface density \textit{in the spiral arm}, which is a few times denser than the average including the interarm region, 
for $l=8$ and $\alpha = 0.35$, we have
\begin{align}\label{roughestimate}
\nonumber
t_c &= \frac{\pi}{4} \left[\frac{(4.302\times 10^{-6})(71 \times 10^6)(0.48)}{(0.48)^2}\right]^{-1/2} \\\nonumber
&= 0.031 \mathrm{ (km/s/kpc)^{-1}} \\
&= 31 \times 10^6 \mathrm{yr},
\end{align}
where we have chosen $\Sigma_g$ to have its half-peak value, 71 $M_\odot$ ${\rm pc}^{-2}$, in the feather.

The rough estimate (\ref{roughestimate}) underestimates the accurate computation of equation (\ref{growthtime}) by a factor of 2.6, suggesting that differential expansion and magnetic stresses in the postshock region play stabilizing influences in the actual feathering phenomenon.  Nevertheless, although the correct mathematical calculation of the feathering instability is complex, the basic mechanism behind its operation is roughly quasi 1-D contraction along magnetic field lines roughly parallel to the spiral arm while the background flow is swept downstream roughly perpendicular to the spiral arm.  Numerical simulation then demonstrates that the full nonlinear development of the instability results, not in permanent collapse for most of the gas in the over-dense regions, but to a redispersal in between the spiral arms as the expanding set of background magnetic fields helps to tear apart the dense condensations that might in a nonmagnetic context have experienced overall continued gravitational collapse. \cite{2007ApJ...662L..75S} suggest that this is the reason why OB star formation, as prominently as they seem to delineate spiral structure and substructure, is actually quite inefficient in its operation in the present, relatively strongly magnetized, universe of interstellar media.  This mental construct, coupled with the visual display of Figure \ref{FIG_VEL}, suggests that giant molecular cloud associations are, not permanent material entities, but a manifestation of 
the parasitic formation and dissolution of the feathers at the crests of the nonlinear density waves that we call gaseous spiral arms.  With this point of view (which we recognize will not be universally accepted), the pattern is long-lived; the individual condensations are not.

\acknowledgments
This research is part of the PhD Thesis Dissertation of WKL in the Physics Department of UCSD. The authors would also like to acknowledge the support of the National Science Council (NSC) of Taiwan for its support of the Theoretical Institute for Advanced Research in Astrophysics (TIARA) based in Academia Sinica's Institute of Astronomy and Astrophysics (ASIAA).  

\appendix
\section{WKBJ APPROXIMATION OF SELF-GRAVITY}
\label{appendixA}
The self-gravity of the gas is governed by the Poisson equation in the thin-disk geometry, 
\begin{equation}
\nabla^2 \mathcal{ V_{\rm g}} = 4\pi G\Sigma\delta(z),
\end{equation}
where $\mathcal{V_{\rm g}}$ and $\Sigma$ are the gravitational potential and surface density of the gas, respectively. We use the above equation for both quasi-1D spiral shock and the 2D feathering perturbation. In the asymptotic approximation (quasi-rectangular) of the spiral coordinates introduced in this paper, we can write the Laplacian in the following form:
\begin{equation}
\nabla^2 = \left(\frac{\varpi \sin{i}}{m}\right)^{-2}\left(\frac{\partial^2}{\partial \eta^2} + \frac{\partial^2}{\partial \xi^2} + \frac{\partial^2}{\partial \zeta^2}\right),
\end{equation}
where $\zeta \equiv z/[(\varpi \sin{i}/m)]$ and $z$ is the physical coordinate perpendicular to the plane of the razor-thin disk. Consistent with the equation (\ref{totalpotential}), we define $\mathcal{V}_{\rm g} \equiv 2\pi G\Sigma_0 (\varpi \sin{i}/m)\phi$, so that $\phi$ is the dimensionless gaseous self-gravitational potential. The dimensionless Poisson equation now reads,
\begin{equation}
\left(\frac{\partial^2}{\partial \eta^2} + \frac{\partial^2}{\partial \xi^2} + \frac{\partial^2}{\partial \zeta^2}\right)\phi = 2 \sigma \delta(\zeta),
\end{equation}
and is subject to the following boundary conditions: $\partial_\zeta \phi |^{+\epsilon}_{-\epsilon} = 2 \sigma$ as the integrability condition across the midplane $\zeta=0$; $\phi\rightarrow 0$ when $\zeta \rightarrow \pm \infty$; periodic boundary conditions for $\eta$ and $\xi$ directions. Note that both $\phi$ and $\partial_\eta \phi$ are continuous across the spiral shock, but the term $\partial^2_\eta \phi$ requires special treatment because of the delta function on the right-hand side. If we Fourier transform in $\eta$ and $\xi$, we have $[-n^2-(l/\tilde{L})^2+\partial^2_\zeta]\tilde{\phi}_{n,l} = 2\tilde{\sigma}_{n,l}\delta(\zeta)$, where $n$ and $l/\tilde{L}$ are the corresponding wavenumbers for these directions. By integrating the last expression across $\zeta = 0$ and requiring exponentially decaying solutions in the $\zeta$ direction for both positive and negative values of $\zeta$, we get the familiar WKBJ result:
\begin{equation}\label{eq1}
\tilde{\phi}_{n,l} = -\frac{\tilde{\sigma}_{n,l}}{\sqrt{n^2+(l/\tilde{L})^2}}.
\end{equation}
For a 1-D TASS density profile, we take $l=0$ and get $\tilde{\phi}_{n,0} = -2\tilde{\sigma}_{n,0}/|n|$. Then it is straight forward to obtain the equations (\ref{sg1}) and (\ref{sg2}) by considering the real and imaginary parts of $\tilde{\sigma}_n$.

For the 2-D feathering perturbation, we adopt a simplification of the complete treatment. For feathering perturbations reminiscent of the substructures observed in real spiral galaxies, $l/\tilde L$ is appreciably larger than the $n$ values needed to reconstruct a reasonable accurate surface density profile of the TASS background state.  To be sure, a formally infinite number of $n$'s are required if we wish to recover the sharp jump of the background shockfront, but this is a feature of the background state and not of the smoother perturbations that we are ascribing to the feathering instability.  Correcting for the finite thickness of the disk would also lead to smoother relations between the perturbed self-gravitational potential and the perturbational surface density.  With the assumption that $l/\tilde L$ is much larger than the $n$ in any of the important Fourier coefficients $\tilde \phi_{n,l}$ and $\tilde{\sigma}_{n,l}$, we adopt the following approximation:
\begin{equation}\label{extremeWKBJ}
\tilde{\phi}_l(\eta) = -\frac{1}{|l/\tilde{L}|} \tilde{\sigma}_l(\eta) \quad
\text{and} \quad \tilde{\phi}'_l(\eta) = -\frac{1}{|l/\tilde{L}|}\tilde{\sigma}'_l(\eta).
\end{equation}
In the actual example shown in Figure \ref{FIG_VEL}, the validity of the approximation is questionable, as it amounts to the assumption that $(l/\tilde L)^2=2^2$ is a large number.  But the feathering displayed in that figure also has too large a spacing between condensations; better examples will be given in Paper II where the approximation used in equation (\ref{extremeWKBJ}) has more justification.

\section{PERTURBATION ON THE LORENTZ FORCE AND INDUCTION EQUATION}
\label{appendixB}
Because the interstellar magnetic field is free of monopoles, it is derivable as the curl of a vector potential $\bf A$.  For a field that lies entirely in a plane, the vector potential can have a single component, ${\bf A} = A(\varpi,\varphi,t)\hat{\mathbf{e}}_z$.  Under these circumstances,
\begin{equation}
\mathbf{B}=\nabla \times (A\hat{\mathbf{e}}_z)=-\hat{\mathbf{e}}_z\times\nabla A.
\end{equation}
The equation for field freezing can now be written,
\begin{equation}\label{fluxfreezing}
\nabla \times \left[ {\partial A\over \partial t}\hat{\mathbf{e}}_z -\left(\hat{\mathbf{e}}_z\times \nabla A\right)\times {\bf u}\right]=0.
\end{equation}
With a proper choice of gauge (namely an initial time-independent state in which $\bf B$ is parallel to $\bf u$, or $\nabla A$ is perpendicular to $\bf u$), we can "uncurl" the above equation, expand the triple vector product, and derive an evolutionary equation for $A$:
\begin{equation}\label{evolutionaryA}
{\partial A\over \partial t} +{\bf u}\cdot \nabla A = 0.
\end{equation}
In other words, field freezing in this context is simply the statement of the conservation of $A$ as we follow the motion of fluid elements.  

If we write $A = A_0 + A_{\rm TASS} + A_1$, and ${\bf u} = {\bf u}_0 + {\bf u}_{\rm TASS} + {\bf u}_1$, the satisfaction of the condition of field freezing by the zeroth order axisymmetric and TASS states implies to linear order that
\begin{equation}\label{pertA}
{\partial A_1 \over \partial t} + \left({\bf u}_0 +{\bf u}_{\rm TASS}\right)\cdot \nabla A_1 = -\left[\nabla\left( A_0 + A_{\rm TASS}\right)\right]\cdot{\bf u}_1.
\end{equation}
The elimination of one spatial derivative in the evolutionary equation for $A_1$ (or $\tilde{A}_1$) explains why there is no equation for $d\tilde{A}_1^\prime/d\eta$ in \S \ref{mos}.  

If we use tilde to denote dimensionless perturbational variables,
\begin{equation}\label{mdef}
\tilde{A}_1 \equiv \frac{m}{\varpi \sin{i}}\frac{A_1}{B_{\xi0}},
\end{equation}
we then get from equation (\ref{pertA}): 
\begin{align}\label{pertA2}
\frac{\varpi \sin{i}}{m}\pd{\tilde{A}_1}{t} + u_\eta \pd{\tilde{A}_1}{\eta} + u_{\xi}\pd{\tilde{A}_1}{\xi} = \frac{\varpi \sin{i}}{m}\left( B_{\xi}u_{\eta 1} - B_{\eta}u_{\xi 1}\right).
\end{align}

Again, we have used unscripted variables to denote the axisymmetric state (denoted with a zero) plus the TASS value (denoted with a hat), and a subscript 1 to denote the dimensioned quantity associated with the feathering perturbations (when nondimensionalized, these are given a tilde).  Using the definition of $\tilde{A}_1$ in equation (\ref{mdef}) and dividing equation (\ref{pertA2}) by $\sqrt{2UV}$, we obtain the dimensionless induction equation:
\begin{align}
\frac{1}{\kappa}\pd{\tilde{A}_1}{t} + (-\nu+\uhat) \pd{\tilde{A}_1}{\eta} + \left(\frac{-\nu}{\tan{i}}+\frac{\kappa}{2\Omega}\vhat\right)\pd{\tilde{A}_1}{\xi} = (1+\shat)\utilde - \left(\frac{\kappa}{2\Omega}\tan{i}\right)\vtilde,
\end{align}
where we have used $\utilde \equiv u_{\eta 1}/\sqrt{2UV}$, $\vtilde \equiv u_{\xi 1}/V$, $\hat{B}_\xi/B_{\xi0}=(1+\shat)$ and $\hat{B}_\eta / B_{\xi 0} = \tan{i}$. In practice, we also need the governing equation of $\tilde{A}_1^\prime \equiv\partial \tilde{A}_1/\partial \eta$, which can be obtained by taking the $\eta$-derivative of equation (\ref{f24}):
\begin{align}\label{f24a}
\frac{1}{\kappa} \pd{\tilde{A}_1^\prime}{t}
+ (-\nu+\uhat) \pd{\tilde{A}_1^\prime}{\eta}+ \od{\uhat}{\eta}\tilde{A}_1^\prime
+ \frac{-\nu}{\tan{i}}\pd{\tilde{A}_1^\prime}{\xi}+\frac{\kappa}{2\Omega} \od{\vhat}{\eta}\pd{\tilde{A}_1^\prime}{\xi}
=(1+\shat)\pd{\utilde}{\eta}+\od{\shat}{\eta}\utilde- \left(\frac{\kappa}{2\Omega}\tan{i}\right)\pd{\vtilde}{\eta}.
\end{align}

To compute the linearized Lorentz force, we use the expression
\begin{equation}
\mathbf{f} = -\frac{2z_0}{4\pi\Sigma}\left[\left(\nabla^2 A\right)\nabla A_1 + \left( \nabla^2 A_1\right) \nabla A \right].
\end{equation}
Again, we have used $A$ as a short hand for $A_0 + A_{\rm TASS}$.  We also ignore the linearized contribution that comes from expanding $\Sigma = \Sigma_0+\Sigma_{\rm TASS}+\Sigma_1$ in the denominator because we are interested the feathering effect in the postshock region, where the perturbation surface density $\Sigma_1$ is very small relative to the axisymmetric and TASS contributions. More explicitly, then, we have to lowest asymptotic order for small $\sin i$:
\begin{equation}\nonumber
\nabla \left( A_0+A_{\rm TASS}\right) \simeq \frac{m}{\varpi\sin{i}}\left(\pd{A}{\eta}\hat{\bf e}_\eta +\pd{A}{\xi}\hat{\bf e}_\xi \right) = -B_\xi\hat{\bf e}_\eta +B_\eta\hat{\bf e}_\xi,
\end{equation}
and,
\begin{align}\nonumber
\nabla^2 \left( A_0+A_{\rm TASS}\right) &\simeq \frac{m}{\varpi\sin{i}}\left(-\pd{B_\xi}{\eta}+\pd{B_\eta}{\xi}\right)=-\frac{m}{\varpi\sin{i}}\od{\hat{B}_\xi}{\eta},
\end{align}
where on the right-hand-sides we have used $A = A_0+A_{\rm TASS}$ as a shorthand. Therefore, the perturbed Lorentz force per unit mass can be written as,
\begin{align}\label{PLF}
\mathbf{f} = -\frac{2z_0}{4\pi\Sigma}\left(\nabla^2 A_1\right)\nabla A -\frac{2z_0}{4\pi\Sigma}\left(\nabla^2 A\right)\nabla A_1,
\end{align}
where we define the perturbed magnetic field, $\mathbf{B}_1=-\hat{\mathbf{e}}_z\times\nabla A_1 $.  For notational convenience, we write the perturbation Lorentz acceleration as coming from two parts: $\mathbf{f}_1 =\mathbf{f}^{(1)}+\mathbf{f}^{(2)}$, where 
\begin{align}\nonumber
\mathbf{f}^{(1)} &= \frac{2z_0}{4\pi\Sigma}\left(\nabla^2 A_1\right)\left(\hat{B}_\xi \hat{\mathbf{e}}_\eta - \hat{B}_\eta \hat{\mathbf{e}}_\xi\right)\\
&\simeq  \frac{m}{\varpi\sin{i}}v^2_{A0}\left(\frac{\partial^2 \tilde{A}_1}{\partial\eta^2} + \frac{\partial^2 \tilde{A}_1}{\partial\xi^2} \right) \left(\hat{\mathbf{e}}_\eta -\frac{\hat{u}_\eta}{\hat{u}_\xi} \hat{\mathbf{e}}_\xi\right),
\end{align}
and,
\begin{align}\nonumber
\mathbf{f}^{(2)} &\simeq \frac{2z_0}{4\pi\Sigma}\left(\frac{m}{\varpi\sin{i}}\right)^2\od{\hat{B}_{\xi}}{\eta}\left(\pd{A_1}{\eta} \hat{\mathbf{e}}_\eta + \pd{A_1}{\xi}\hat{\mathbf{e}}_\xi\right) \\
&\simeq \frac{m}{\varpi\sin{i}}\frac{v^2_{\rm A0}}{1+\shat}\od{\shat}{\eta}\left(\pd{\tilde{A}_1}{\eta} \hat{\mathbf{e}}_\eta + \pd{\tilde{A}_1}{\xi}\hat{\mathbf{e}}_\xi\right).
\end{align}
In the above, we have approximated $\cos^2{i} \simeq 1$ for small $\sin i$. The dimensionless components of the Lorentz force $({f}_\eta, {f}_\xi)$ can be found by rearranging the terms:
\begin{align}\nonumber
{f}_\eta =& \left(\frac{\varpi\sin{i}/m}{2UV}\right) {\rm f}_{1\eta} \\\nonumber
=& x_{\rm A0}\left(\frac{\partial^2}{\partial\eta^2}+\frac{\partial^2}{\partial\xi^2} \right) \tilde{A}_1 +x_{\rm A0}\frac{\shat^\prime}{1+\shat}\pd{\tilde{A}_1}{\eta}
\end{align}
and,
\begin{align}
\nonumber
{f}_\xi =&\left(\frac{\varpi\sin{i}/m}{V\sqrt{2UV}}\right){\rm f}_{1\xi}\\\nonumber
=& -\frac{2\Omega}{\kappa}x_{\rm A0}\left(\frac{\tan{i}}{1+\shat}\right)\left(\frac{\partial^2}{\partial\eta^2} + \frac{\partial^2}{\partial\xi^2} \right)\tilde{A}_1+\frac{2\Omega}{\kappa}x_{\rm A0}\frac{\shat'}{1+\shat}\frac{\partial\tilde{A}_1}{\partial\xi} ,
\end{align}
where the factor $(\varpi\sin{i}/m)$ is included for consistency with the convention in our dimensionless variables (which will be eventually cancelled). 

\section{MATRICES}
Here we list out the coefficient matrices for the ODEs (see, eqs \ref{s01}, \ref{s02}, \ref{s03} and \ref{s04}) involved in the calculation (in the form of $\mathbf{A}(\eta)\mathbf{V}^\prime(\eta) = \mathbf{B}(\eta)\mathbf{V}(\eta)$) before the reduction procedure in the \S \ref{mos}. The \textit{four} ODEs and the \textit{five} tilde variables corresponds to the columns and rows, respectively. We obtain the square matrices by eliminating the fifth column with the use of induction equation (see text). Basically the matrices are collections of the background terms in the equations and boundary conditions. For the purpose of clarity, we define the following: $u_{\rm T} \equiv -\nu+\uhat$, $\sigma_{\rm T} \equiv 1+\shat$ and $\omega_{\rm T} \equiv \omega - (l/\tilde{L})\hat{v}_{\rm T} $. The ``mass matrix", $\mathbf{\hat{A}}_{\omega,l}$:
\begin{equation}
	\begin{pmatrix}
		u_{\rm T} & \sigma_{\rm T} & 0 & 0 & 0 \\ 
		\hat{b} & u_{\rm T} & 0 & -x_{\rm A0}\shat^\prime/\sigma_{\rm T}  & -x_{\rm A0} \\ 
		0 & 0 & u_{\rm T} & 0 & \hat{h}x_{\rm A0}\\
		0 & 0 & 0 & u_{\rm T} & 0 \\
	\end{pmatrix},
\end{equation}
where
\begin{equation}
	\hat{b} \equiv \frac{x_{\rm t0}}{(1+\shat)^2}-\alpha\frac{\tilde{L}}{|l|} \quad \text{and} \quad
	\hat{h} \equiv \frac{2\Omega}{\kappa}\left(\frac{-\nu+\uhat}{-\nu/\tan{i}}\right)=\frac{2\Omega}{\kappa}\frac{\tan{i}}{1+\shat},
\end{equation}
The matrix $\mathbf{\hat{B}}_{\omega,l}$ is given by
\begin{equation}
	\begin{pmatrix}
		-\uhat^\prime-i\omega_{\rm T} & -\shat^\prime & i(l/\tilde{L})(\kappa/2\Omega)\sigma_{\rm T} & 0 & 0 \\
		2x_{\rm T0}\shat^\prime/\sigma_{\rm T}^3 & -\uhat^\prime-i\omega_{\rm T} & 1 & -x_{\rm A0}(l/\tilde{L})^2 & 0\\
		(2\Omega/\kappa)(il/\tilde{L})\hat{b} & -(1+\vhat') & -i\omega_{\rm T} & (2\Omega/\kappa)x_{\rm A0}/\sigma_{\rm T}\left[\tan{i}(l/\tilde{L})^2-\shat'(il/\tilde{L})\right] & 0  \\
		0 & \sigma_{\rm T} & -(\kappa/2\Omega)\tan{i} & -i\omega_T & 0 \\
	\end{pmatrix}.
\end{equation}

\bibliography{Feathering}

\end{document}

%% file: symbols.tex
\newcommand{\od}[2]{ \frac{d #1}{d #2} }
\newcommand{\pd}[2]{{\frac{\partial #1}{\partial #2}}}

\newcommand{\uhat}{\hat{u}}
\newcommand{\vhat}{\hat{v}}
\newcommand{\shat}{\hat{\sigma}}
\newcommand{\xhat}{\hat{x}}
\newcommand{\phat}{\hat{\phi}}

\newcommand{\utilde}{\tilde{u}}
\newcommand{\vtilde}{\tilde{v}}
\newcommand{\stilde}{\tilde{\sigma}}
\newcommand{\ptilde}{\tilde{\phi}}

\newcommand{\utwl}{\tilde{u}_{\omega,l}}
\newcommand{\vtwl}{\tilde{v}_{\omega,l}}
\newcommand{\stwl}{\tilde{\sigma}_{\omega,l}}
\newcommand{\ptwl}{\tilde{\phi}_{\omega,l}}


%% file: ver9.bbl
\begin{thebibliography}{36}
\expandafter\ifx\csname natexlab\endcsname\relax\def\natexlab#1{#1}\fi

\bibitem[{{Ascher} {et~al.}(1995){Ascher}, {Mattheij}, \&
  {Russell}}]{ascher1995numerical}
{Ascher}, U.~M., {Mattheij}, R., \& {Russell}, R.~D. 1995, Numerical solution
  of boundary value problems for ordinary differential equations (Philadelphia:
  Society for Industrial and Applied Mathematics)

\bibitem[{{Balbus}(1988)}]{B88}
{Balbus}, S.~A. 1988, \apj, 324, 60

\bibitem[{{Block} {et~al.}(1994){Block}, {Bertin}, {Stockton}, {Grosbol},
  {Moorwood}, \& {Peletier}}]{1994A&A...288..365B}
{Block}, D.~L., {Bertin}, G., {Stockton}, A., {Grosbol}, P., {Moorwood},
  A.~F.~M., \& {Peletier}, R.~F. 1994, \aap, 288, 365

\bibitem[{{Block} {et~al.}(1996){Block}, {Elmegreen}, \&
  {Wainscoat}}]{1996Natur.381..674B}
{Block}, D.~L., {Elmegreen}, B.~G., \& {Wainscoat}, R.~J. 1996, \nat, 381, 674

\bibitem[{{Block} \& {Wainscoat}(1991)}]{1991Natur.353...48B}
{Block}, D.~L., \& {Wainscoat}, R.~J. 1991, \nat, 353, 48

\bibitem[{{Chakrabarti} {et~al.}(2003){Chakrabarti}, {Laughlin}, \&
  {Shu}}]{2003ApJ...596..220C}
{Chakrabarti}, S., {Laughlin}, G., \& {Shu}, F.~H. 2003, \apj, 596, 220

\bibitem[{{Corder} {et~al.}(2008){Corder}, {Sheth}, {Scoville}, {Koda},
  {Vogel}, \& {Ostriker}}]{2008ApJ...689..148C}
{Corder}, S., {Sheth}, K., {Scoville}, N.~Z., {Koda}, J., {Vogel}, S.~N., \&
  {Ostriker}, E. 2008, \apj, 689, 148

\bibitem[{{Dobbs}(2008)}]{2008MNRAS.391..844D}
{Dobbs}, C.~L. 2008, \mnras, 391, 844

\bibitem[{{Dobbs} \& {Bonnell}(2006)}]{2006MNRAS.367..873D}
{Dobbs}, C.~L., \& {Bonnell}, I.~A. 2006, \mnras, 367, 873

\bibitem[{{D'Onghia} {et~al.}(2012){D'Onghia}, {Vogelsberger}, \&
  {Hernquist}}]{2012arXiv1204.0513D}
{D'Onghia}, E., {Vogelsberger}, M., \& {Hernquist}, L. 2012, ArXiv e-prints

\bibitem[{{Dwarkadas} \& {Balbus}(1996)}]{1996ApJ...467...87D}
{Dwarkadas}, V.~V., \& {Balbus}, S.~A. 1996, \apj, 467, 87

\bibitem[{{Elmegreen}(1980)}]{1980ApJ...242..528E}
{Elmegreen}, D.~M. 1980, \apj, 242, 528

\bibitem[{{Goldreich} \& {Lynden-Bell}(1965)}]{1965MNRAS.130..125G}
{Goldreich}, P., \& {Lynden-Bell}, D. 1965, \mnras, 130, 125

\bibitem[{{Julian} \& {Toomre}(1966)}]{1966ApJ...146..810J}
{Julian}, W.~H., \& {Toomre}, A. 1966, \apj, 146, 810

\bibitem[{{Kendall} {et~al.}(2008){Kendall}, {Kennicutt}, {Clarke}, \&
  {Thornley}}]{2008MNRAS.387.1007K}
{Kendall}, S., {Kennicutt}, R.~C., {Clarke}, C., \& {Thornley}, M.~D. 2008,
  \mnras, 387, 1007

\bibitem[{{Kim} \& {Ostriker}(2002)}]{KO2002}
{Kim}, W., \& {Ostriker}, E.~C. 2002, \apj, 570, 132

\bibitem[{{Kim} \& {Ostriker}(2006)}]{2006ApJ...646..213K}
{Kim}, W.-T., \& {Ostriker}, E.~C. 2006, \apj, 646, 213

\bibitem[{{La Vigne} {et~al.}(2006){La Vigne}, {Vogel}, \&
  {Ostriker}}]{2006ApJ...650..818L}
{La Vigne}, M.~A., {Vogel}, S.~N., \& {Ostriker}, E.~C. 2006, \apj, 650, 818

\bibitem[{{Lin} \& {Shu}(1964)}]{1964ApJ...140..646L}
{Lin}, C.~C., \& {Shu}, F.~H. 1964, \apj, 140, 646

\bibitem[{{Lizano} \& {Shu}(1989)}]{1989ApJ...342..834L}
{Lizano}, S., \& {Shu}, F.~H. 1989, \apj, 342, 834

\bibitem[{{Lubow} {et~al.}(1986){Lubow}, {Cowie}, \&
  {Balbus}}]{1986ApJ...309..496L}
{Lubow}, S.~H., {Cowie}, L.~L., \& {Balbus}, S.~A. 1986, \apj, 309, 496

\bibitem[{{Lynds}(1970)}]{1970IAUS...38...26L}
{Lynds}, B.~T. 1970, in IAU Symposium, Vol.~38, The Spiral Structure of our
  Galaxy, ed. {W.~Becker \& G.~I.~Kontopoulos}, 26

\bibitem[{{Mathewson} {et~al.}(1972){Mathewson}, {van der Kruit}, \&
  {Brouw}}]{1972A&A....17..468M}
{Mathewson}, D.~S., {van der Kruit}, P.~C., \& {Brouw}, W.~N. 1972, \aap, 17,
  468

\bibitem[{{Mouschovias} {et~al.}(1974){Mouschovias}, {Shu}, \&
  {Woodward}}]{1974A&A....33...73M}
{Mouschovias}, T.~C., {Shu}, F.~H., \& {Woodward}, P.~R. 1974, \aap, 33, 73

\bibitem[{{Parker}(1969)}]{1969SSRv....9..651P}
{Parker}, E.~N. 1969, \ssr, 9, 651

\bibitem[{{Piddington}(1973)}]{1973ApJ...179..755P}
{Piddington}, J.~H. 1973, \apj, 179, 755

\bibitem[{{Roberts}(1969)}]{Roberts1969a}
{Roberts}, W.~W. 1969, \apj, 158, 123

\bibitem[{{Roberts} \& {Yuan}(1970)}]{1970ApJ...161..887R}
{Roberts}, Jr., W.~W., \& {Yuan}, C. 1970, \apj, 161, 887

\bibitem[{{Scoville} \& {Rector}(2001)}]{Scoville2001press}
{Scoville}, N., \& {Rector}, T. 2001, HST press release

\bibitem[{{Sellwood}(2012)}]{2012ApJ...751...44S}
{Sellwood}, J.~A. 2012, \apj, 751, 44

\bibitem[{{Shetty} \& {Ostriker}(2006)}]{2006ApJ...647..997S}
{Shetty}, R., \& {Ostriker}, E.~C. 2006, \apj, 647, 997

\bibitem[{{Shu} {et~al.}(2007){Shu}, {Allen}, {Lizano}, \&
  {Galli}}]{2007ApJ...662L..75S}
{Shu}, F.~H., {Allen}, R.~J., {Lizano}, S., \& {Galli}, D. 2007, \apjl, 662,
  L75

\bibitem[{Shu {et~al.}(1973)Shu, Milione, \& Roberts}]{SMR1973}
Shu, F.~H., Milione, V., \& Roberts, W.~W. 1973, Astrophysical Journal, 183,
  819

\bibitem[{{Sofue} {et~al.}(1999){Sofue}, {Tutui}, {Honma}, {Tomita},
  {Takamiya}, {Koda}, \& {Takeda}}]{1999ApJ...523..136S}
{Sofue}, Y., {Tutui}, Y., {Honma}, M., {Tomita}, A., {Takamiya}, T., {Koda},
  J., \& {Takeda}, Y. 1999, \apj, 523, 136

\bibitem[{{Spitzer}(1968)}]{1968nim..book....1S}
{Spitzer}, Jr., L. 1968, {Dynamics of Interstellar Matter and the Formation of
  Stars}, ed. B.~M. {Middlehurst} \& L.~H. {Aller} (the University of Chicago
  Press), 1

\bibitem[{{Westpfahl}(1998)}]{1998ApJS..115..203W}
{Westpfahl}, D.~J. 1998, \apjs, 115, 203

\end{thebibliography}
